\documentclass[]{fairmeta_gr2}
\usepackage{multirow}
\usepackage{color}
\usepackage{colortbl}
\usepackage{mdframed}
\usepackage{caption}
\usepackage{subcaption}
\usepackage{algpseudocode}
\usepackage{algorithm}
\usepackage{amsthm}
\usepackage{amssymb}
\usepackage{graphicx}
\usepackage[normalem]{ulem}

\usepackage{booktabs}
\newcommand{\cmark}{\checkmark}

\usepackage[most]{tcolorbox}

\title{GR2: Generative Reasoning Re-ranker}
\author[1]{Mingfu Liang}
\author[1]{Yufei Li}
\author[1]{Jay Xu}
\author[]{Kavosh Asadi}
\author[2]{Xi Liu}
\author[]{Shuo Gu}
\author[]{Kaushik Rangadurai}
\author[]{Frank Shyu}
\author[]{Shuaiwen Wang}
\author[]{Song Yang}
\author[]{Zhijing Li}
\author[]{Jiang Liu}
\author[]{Mengying Sun}
\author[]{Fei Tian}
\author[]{Xiaohan Wei}
\author[]{Chonglin Sun}
\author[]{Jacob Tao}
\author[]{Shike Mei}
\author[]{Wenlin Chen}
\author[]{Santanu Kolay}
\author[]{Sandeep Pandey}
\author[3]{Hamed Firooz}
\author[3]{Luke Simon}

\affiliation[]{Meta AI}
\contribution[1]{Joint First Author}
\contribution[2]{Project Lead}
\contribution[3]{Direction Lead}

\abstract{Recent studies increasingly explore Large Language Models (LLMs) as a new paradigm for recommendation systems due to their scalability and world knowledge. However, existing work has three key limitations: (1) most efforts focus on retrieval and ranking, while the reranking phase, critical for refining final recommendations, is largely overlooked; (2) LLMs are typically used in zero-shot or supervised fine-tuning settings, leaving their reasoning abilities, especially those enhanced through reinforcement learning (RL) and high-quality reasoning data, underexploited; (3) items are commonly represented by non-semantic IDs, creating major scalability challenges in industrial systems with billions of identifiers. To address these gaps, we propose the Generative Reasoning Reranker (GR2), an end-to-end framework with a three-stage training pipeline tailored for reranking. First, a pretrained LLM is mid-trained on semantic IDs encoded from non-semantic IDs via a tokenizer achieving $\ge$99\% uniqueness. Next, a stronger larger-scale LLM generates high-quality reasoning traces through carefully designed prompting and rejection sampling, which are used for supervised fine-tuning to impart foundational reasoning skills. Finally, we apply Decoupled Clip and Dynamic sAmpling Policy Optimization (DAPO), enabling scalable RL supervision with verifiable rewards designed specifically for reranking. Experiments on two real-world datasets demonstrate GR2's effectiveness: it surpasses the state-of-the-art OneRec-Think by 2.4\% in Recall@5 and 1.3\% in NDCG@5. Ablations confirm that advanced reasoning traces yield substantial gains across metrics. We further find that RL reward design is crucial in reranking: LLMs tend to exploit reward hacking by preserving item order, motivating conditional verifiable rewards to mitigate this behavior and optimize reranking performance.}

\date{January 1st, 2026}

\begin{document}

\maketitle

\section{Introduction}
Recommendation systems have become indispensable across a wide spectrum of online platforms, serving to mitigate information overload by intelligently identifying and presenting items aligned with users’ interests and needs~\citep{ramanujam2025large,behdin2025scaling,deng2025onerec}. Over the past decade, deep neural networks have been extensively adopted to model the complex relationships between user feedback and vast arrays of item and user features, typically encoded via large-scale embedding tables~\citep{liang2025external,luo2025meta,zhang2024wukong,zhang2022dhen,wang2021dcn, zhou2019deep}.
Recently, Large Language Models (LLMs) have emerged as a transformative paradigm for recommendation systems~\citep{zhou2025efficiency,zhang2025reasonrec, zhao2024recommender,wu2024survey}, driven by their remarkable capacity for continual performance improvement through model scaling, comprehensive world knowledge, and nuanced contextual understanding. Notable examples include P5~\citep{geng2022recommendation}, which unifies diverse recommendation tasks within a single LLM model, OneRec-Think~\citep{liu2025onerec}, which integrates retrieval and ranking stages by fine-tuning LLMs, and PLUM~\citep{he2025plum} that adapts pre-trained large language models to deliver scalable and efficient recommendations for YouTube, enhancing both retrieval quality and system performance.

Re-ranking is directly responsible for refining and enhancing recommendation outcome and thus a critical component of modern recommender systems~\citep{gao2025llm4rerank,gao2024smlp4rec,liuneural}. Despite its significance, the re-ranking stage is frequently neglected in recent LLM-based approaches. 
Furthermore, these studies often deploy LLMs in zero-shot scenarios or fine-tune them on recommendation datasets without reasoning trace, thereby underutilizing the models’ reasoning capabilities—particularly those that can be enhanced through reinforcement learning (RL) and the incorporation of high-quality reasoning data. 
Another prevailing limitation is the reliance on non-semantic item identifiers, which poses substantial scalability and adaptability challenges in industrial environments, where the sheer volume of item IDs can lead to an unmanageable expansion of the LLM’s vocabulary.

\begin{figure}[htbp]
  \centering
  \includegraphics[width=\linewidth]{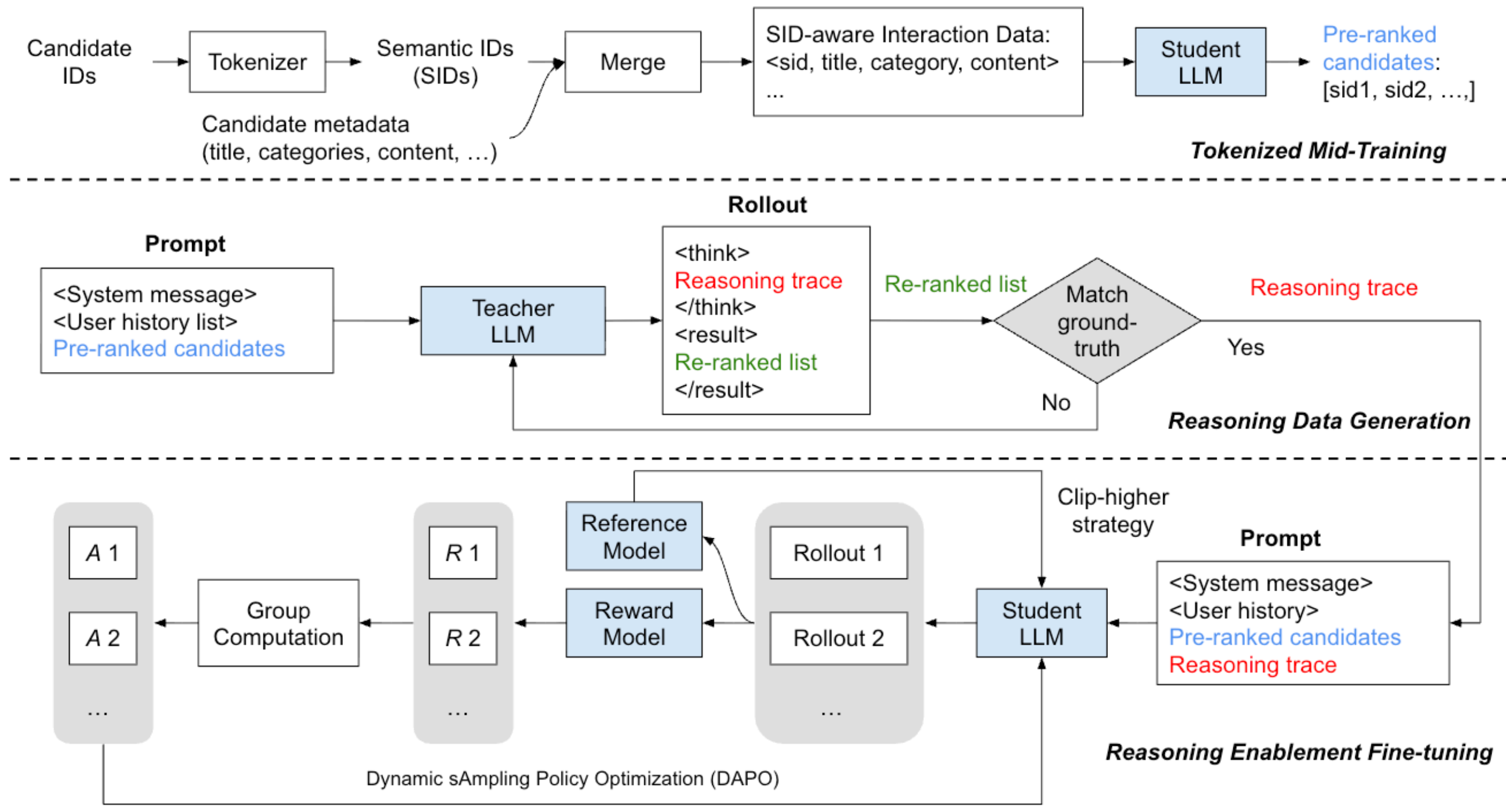}
  \caption{Overview of the 3-stage training pipeline: student LLM mid-training on tokenized semantic IDs (up), reasoning data generation with teacher LLM and rejection sampling (middle), and student LLM reasoning enablement by SFT and RL (down)}
  \label{fig:framework}
\end{figure}

To address the aforementioned limitations, we introduce the Generative Reasoning Re-ranker (GR2), an advanced LLM-based recommendation framework that bridges semantic item representations with world knowledge and sophisticated reasoning to effectively re-rank candidate lists. 
GR2 is architected around a three-stage training pipeline, as illustrated in Figure~\ref{fig:framework}, meticulously tailored for the demands of re-ranking in large-scale recommendation systems. 
In the first stage, a pre-trained student LLM (e.g., Qwen3-8B~\citep{qwen3technicalreport}) undergoes mid-training on semantic item identifiers inspired by OneRec-Think~\citep{liu2025onerec}. 
These semantic IDs are derived from non-semantic raw item IDs using a tokenizer designed to achieve at least 99\% uniqueness, enabling the model to better capture item semantics, generalize robustly across diverse item spaces, and ensure reliable item distinguishability.
The second stage leverages a more powerful teacher LLM (e.g., Qwen3-32B~\citep{qwen3technicalreport}) with larger model size to generate high-quality, hierarchical reasoning traces. 
This is accomplished through prompts carefully crafted for re-ranking problem and rigorous rejection sampling, ensuring that the generated traces are both relevant and informative. 
These reasoning traces serve as the foundation for supervised fine-tuning, equipping the model with basic reasoning skills and enabling it to interpret and connect item semantics with user preferences.
In the final stage, GR2 adapts DAPO~\citep{yu2025dapo} by re-designing its reward function explicitly for the re-ranking task. 
The adapted DAPO provides scalable supervision through a custom reward function that aligns with the objectives of re-ranking, further refining the student LLM’s reasoning capabilities and enhancing its ability to deliver highly relevant and personalized re-ranked candidate list.
The main contributions of the paper are summarized as follows
\begin{itemize}
\item Distinct from prior work that primarily adapts LLMs for retrieval and late-stage ranking, we investigate and establish design principles for LLMs tailored to the re-ranking stage in recommendation systems. 

\item We introduce a robust method for transforming non-semantic item identifiers into semantic IDs, achieving $\geq99\%$ uniqueness. 
Through mid-training on a data mixture of semantic IDs and world knowledge, we enable the LLM to recognize and reason over these enriched representations, effectively bridging item semantics with external knowledge.

\item Recognizing the centrality of reasoning in re-ranking, we develop specialized prompts tailored for the re-ranking context to elicit high-quality, hierarchical reasoning traces. 
Rejection sampling is employed to filter out noisy traces, and the model is subsequently fine-tuned on these curated traces to impart foundational reasoning capabilities.

\item To further augment the model’s reasoning proficiency, we adapt the DAPO algorithm with a custom reward function specifically designed for re-ranking scenarios. 
This stage provides scalable, reward-driven supervision, enabling the LLM to refine its reasoning process and deliver superior re-ranking performance.

\item Our comprehensive evaluation of GR2 on two real-world datasets demonstrates its superior performance: GR2 consistently surpasses the leading baseline, OneRec-Think, across both Recall@K and NDCG@K metrics. 
Furthermore, ablation studies validate the vital contributions of reasoning and RL, uncovering additional insights into our design.

\end{itemize}

The remainder of this paper is structured as follows. Section~\ref{sec:stage_1} details the first-stage tokenized mid-training process, including the encoding of engagement signals via contrastive loss and techniques for achieving high uniqueness.
Section~\ref{sec:stage_2} elaborates on the generation of reasoning data, such as prompt engineering to foster re-ranking-oriented reasoning and the application of rejection sampling techniques to denoise.
In Section~\ref{sec:stage_3}, we present methods for enabling reasoning capabilities through advanced RL, featuring reward functions specifically tailored for re-ranking tasks.
Section~\ref{sec:related} provides a concise review of related work in LLM-based recommendation systems and LLM reasoning.
Experiments and analyses are carried out in Section~\ref{sec:exp}. 
Finally, Section~\ref{sec:con} concludes the paper, with additional details provided in the Appendix.
\section{Tokenized Mid-Training}\label{sec:stage_1}
In this section, we introduce the key components for attaining the LLM-based generative recommendation model: \textbf{Tokenization} and \textbf{Mid-Training}. 
To improve the cookbook utilization, we introduce our enhancement for the SID generation. 
To obtain the generative retriever, we stemmed from the mid-training workflow from the state-of-the-art LLM-based generative retriever that perform multi-task post-training on an LLM.

\subsection{Tokenizer and Semantic ID (SID)}

The technique known as semantic ID (SID) has been widely adopted in recommendation systems to mitigate the scalability issues of large embedding tables arising from massive item and user vocabularies. 
The seminal work TIGER~\citep{rajput2023recommender} is the first to apply this technique to sequential recommendation. 
At a high level, given the textual feature $x$ of an item, the tokenizer maps it to a sequence of discrete integers, i.e., a compact symbolic representation, defined as
\begin{equation}
\boxed{
\text{Tokenizer}(x) = (z_1, z_2, \ldots, z_K),
\quad z_k \in \{1,\ldots,C_k\}
}
\end{equation}
where $C_i$ is the cardinality of the $i$-th codebook. The core of the tokenizer is a Residual-Quantized Variational Autoencoder (RQ-VAE)~\citep{lee2022autoregressive}. 
Since SID itself is not the primary contribution of this paper, we defer the detailed formulation to Appendix~\ref{sec:appendix sid}.

\subsection{Contrastive Loss} 
\label{sec:contrastive-loss}
An effective way to encode the item co-engagement history is to apply a contrastive loss. 
For the $i$-th anchor item $x_i$, the contrastive loss is presented as:
\begin{equation}
\mathcal{L}_{\mathrm{ctr}}(x_i)
=
-
\frac{1}{|\mathcal{P}(i)|}
\sum_{p \in \mathcal{P}(i)}
\log
\frac{
\exp\!\left( s(x_i, x_p) \right)
}{
\exp\!\left( s(x_i, x_p) \right)
+
\sum_{n \in \mathcal{N}(i)}
\exp\!\left( s(x_i, x_n) \right)
}.
\label{eq:infonce}
\end{equation}
where $\mathcal{P}(i)$ denote the set of indices of co-engaged (positive) samples, $\mathcal{N}(i)$ denote a set of randomly sampled non-co-engaged (negative) samples. 
The similarity between two items $x_i$ and $x_j$ is defined as a temperature ($T > 0$)-scaled cosine similarity:
\begin{align}
s(x_i, x_j)=\frac{\tilde{\mathbf{h}}_i^\top \tilde{\mathbf{h}}_j}{T} ,\quad\text{where}\  \tilde{\mathbf{h}} = \frac{\mathbf{h}}{\|\mathbf{h}\|_2},
\end{align}
and $\mathbf h$ is the dense embedding of the item's textual feature: $\mathbf{h} = f_{\text{enc}}(x)$.

\subsection{Techniques for Balanced Codebook Utilization}
\label{sec:codebook-techniques}

A key challenge in training RQ-VAE~\citep{lee2022autoregressive} is \emph{codebook collapse}, where only a small subset of codebook entries are actively used while the rest remain ``dead.'' 
This leads to poor reconstruction quality and high collision rates in the generated semantic IDs. We employ five complementary techniques to ensure balanced codebook utilization.

\paragraph{RQ-K-means Initialization.}
Rather than randomly initializing codebooks, we apply the k-means++~\citep{arthur2007kmeansplusplus} clustering on the residuals, e.g., RQ-K-Means to initialize the codebooks. 
This is our default setup and is applied in all of our configures.

\paragraph{Exponential Moving Average (EMA) Codebook Updates.}
Instead of updating codebook entries via gradient descent, we use exponential moving averages for updating the $k$-th codebook:
\begin{align}
    \tilde{\mathbf e}^{(k)} &\leftarrow \texttt{Optimizer}(\mathbf e^{(k)}, \nabla_{\mathbf e^{(k)}}\mathcal L)\\
    \mathbf e^{(k)} &\leftarrow \gamma \mathbf e^{(k)} + (1 - \gamma) \tilde{\mathbf e}^{(k)}
\end{align}
where $\gamma \in [0.95, 0.99]$ is the decay rate. EMA updates provide more stable learning for the codebook by avoiding direct gradient-based optimization, which has been proved crucial for a consistent codebook usage~\citep{van2017neural}.

\paragraph{Diversity Loss.}

To encourage uniform usage of codebook entries, we introduce a diversity 
regularization term for each codebook. For the $k$-th codebook with $C_k$ entries, let
\begin{equation}
p_{k,j} = \frac{n_{k,j}}{\sum_{m=1}^{C_k} n_{k,m}},
\end{equation}
where $n_{k,j}$ denotes the number of assignments in the current 
mini-batch to the $j$-th entry of the $k$-th codebook. 
In practice, $n_{k,j}$ is computed using soft assignment probabilities 
obtained via a softmax over negative distances. The diversity loss is defined as
\begin{equation}
\mathcal{L}_{\text{div}}
=
\lambda_{\text{div}}
\sum_{k=1}^{K}
C_k
\sum_{j=1}^{C_k}
p_{k,j}^2.
\end{equation}

This objective is minimized when 
\(
p_{k,j} = 1/C_k
\)
for all $j$ and all $k$, corresponding to uniform usage of all entries in each codebook.

\paragraph{Dead Code Reset.}
We track the number of consecutive batches each codebook entry has been unused. 
When a code remains unused for more than $\tau$ batches (the reset threshold), we reinitialize it with a ``hard'' sample---one that has the largest reconstruction error under the current codebook. 
Here we omit the codebook level, i.e., the superscript, for brevity. Suppose for the $k$-th codebook, if $\texttt{unused\_count}[k] \geq \tau$
\begin{align}
    \mathbf{e}_k &\leftarrow \mathbf{r}_{i^*}\\
    where\quad 
    i^* &= \arg\max_i \|\mathbf{r}_i - \mathbf{e}_{z_i}\|_2
\end{align}
This mechanism revives dead codes by assigning them to poorly-fitted samples, improving overall codebook coverage.

\paragraph{Random Last Levels.}
To further improve semantic ID uniqueness while preserving semantic meaning in the early levels, we introduce random assignment for the last $M$ levels during inference:
\begin{align}
    k_\ell = \begin{cases}
        \arg\min_k \|\mathbf{r}^{(\ell)} - \mathbf{e}_k^{(\ell)}\|^2 & \text{if } \ell \leq L - M \\
        \texttt{Uniform}(1, K) & \text{if } \ell > L - M
    \end{cases}
\end{align}
The intuition is that early levels capture coarse semantic categories (e.g., product type), while later levels encode finer details. 
By randomizing the last $M$ levels, we trade off some reconstruction fidelity for guaranteed uniqueness of the generated IDs.

\subsection{Mid-Training through Multi-Task Training Strategy}
\label{sec:item alignment}

TIGER~\citep{rajput2023recommender} trains an autoregressive model purely over Semantic ID sequences. In the era of LLMs, leveraging world knowledge encoded in pretrained LMs has become an important direction. 
To this end, OneRec-Think~\citep{liu2025onerec} developed by Kuaishou introduces a mid-training stage (termed as `item alignment' stage) for the pretrained LLM, a key design that enables the LLM to align the recommendation knowledge and language (Semantic IDs) with the LLM's linguistic space and world knowledge.

The key idea of the mid-training stage is to \textbf{interleave Semantic IDs (SIDs) with natural language tokens within a single sequence} and to optimize the SID embedding table through the \textbf{next-token prediction objective}. 
We follow the setup of item alignment from the OneRec-Think~\citep{liu2025onerec} and present  examples of item alignment tasks in Appendix~\ref{sec:item alignment examples}. 
In addition, Google’s PLUM~\citep{he2025plum} incorporates a similar training stage, referred to as \textbf{Continued Pre-training (CPT)}, which follows the same core idea as item alignment.

% Define colors for each principle
\definecolor{principleA}{RGB}{41, 128, 185}   % Blue - System Role
\definecolor{principleB}{RGB}{192, 57, 43}    % Red - Output Constraints
\definecolor{principleC}{RGB}{39, 174, 96}    % Green - Multi-Step Reasoning
\definecolor{principleD}{RGB}{142, 68, 173}   % Purple - Domain Knowledge
\definecolor{principleE}{RGB}{243, 156, 18}   % Orange - Collaborative Context
\definecolor{sidcolor}{RGB}{0, 100, 180}      % SID blue

% Define colors for roles and principles
\definecolor{systemcolor}{RGB}{52, 73, 94}      % Dark gray-blue - System
\definecolor{usercolor}{RGB}{41, 128, 185}      % Blue - User
\definecolor{assistantcolor}{RGB}{39, 174, 96}  % Green - Assistant
\definecolor{sidcolor}{RGB}{155, 89, 182}       % Purple - SID tokens
\definecolor{structcolor}{RGB}{230, 126, 34}    % Orange - Structure
\definecolor{reasoncolor}{RGB}{192, 57, 43}     % Red - Reasoning

\section{Reasoning Data Generation}\label{sec:stage_2}
In this section, we elucidate the design of reasoning data generation for the re-ranking task. 
In Sec.~\ref{Prompt to Incentivize Reasoning to Re-rank}, we introduce a chat-format training sample structure that grounds item representation in semantic IDs and enables chain-of-thought reasoning with structured JSON output. 
In Sec.~\ref{Reasoning Trace Generation}, we present two complementary strategies for generating high-quality reasoning traces—targeted sampling, which leverages ground-truth guidance, and rejection sampling, which ensures reasoning authenticity through iterative verification—along with five prompt design principles that elicit grounded, interpretable, and domain-aware reasoning from large language models.

\subsection{Chat-formatted Template of Reasoning Training Data}
\label{Prompt to Incentivize Reasoning to Re-rank}

To train the LLM for re-ranking with semantic ID reasoning, we construct training samples
in a chat-format structure consisting of three message roles. 
The design follows six key principles:

\begin{itemize}
  \item \textbf{Role-based System Prompt}: The system message establishes an expert
  persona (e.g., ``professional recommendation expert'') and specifies the re-ranking
  objective, activating the model's domain-specific reasoning capabilities.

  \item \textbf{SID-grounded Item Representation}: Both purchase history items and
  candidate items are tagged with semantic ID tokens (e.g.,
  \texttt{<|sid\_begin|><s\_a\_57>...<|sid\_end|>}), enabling the model to learn
  reasoning patterns that naturally interleave semantic IDs with natural language.

  \item \textbf{Rich Item Metadata}: Each item is accompanied by its title and
  category hierarchy, providing multi-modal context that supports both semantic
  understanding and category-aware reasoning.

  \item \textbf{Unified Item Format}: Purchase history and candidate items share
  the same structural format (SID + title + categories), ensuring consistent
  representation learning across input contexts.

  \item \textbf{Chain-of-Thought Reasoning Trace}: The assistant response contains
  step-by-step reasoning, that explicitly cites items by their SIDs, teaching the model
  to produce grounded, verifiable reasoning traces.

  \item \textbf{Structured JSON Output}: The response containing the reasoning trace is formatted as a JSON object
  containing both the reasoning explanation and a ranked recommendation list, enabling
  direct parsing and evaluation of model outputs.
\end{itemize}

%% HIGH-LEVEL STRUCTURE
\begin{tcolorbox}[
    colback=gray!8,
    colframe=gray!60,
    title=\textbf{Chat-Format Training Sample Structure},
    fonttitle=\bfseries,
    boxrule=0.5pt,
    rounded corners,
    breakable,
    enhanced
]

%% SYSTEM MESSAGE
\begin{tcolorbox}[
    colback=systemcolor!10,
    colframe=systemcolor!80,
    title={\small\textbf{System Message} -- Role Definition \& Task Specification},
    fonttitle=\bfseries\color{white},
    colbacktitle=systemcolor,
    boxrule=0.4pt,
    rounded corners,
    left=3pt, right=3pt, top=2pt, bottom=2pt
]
\small\ttfamily
\{\par
\hspace{1em}"role": "system",\par
\hspace{1em}"content": "You are a \textbf{professional e-commerce recommendation expert}  specializing in sequential purchase prediction. YOUR TASK: Predict which item the user is MOST LIKELY TO PURCHASE NEXT by re-ranking 10 pre-ranked candidates from a generative retrieval model."\par
\}\par
\vspace{0.3em}
\rmfamily\color{gray}\textit{$\triangleright$ Establishes expert persona and defines the re-ranking objective}
\end{tcolorbox}

\vspace{0.5em}

%% USER MESSAGE
\begin{tcolorbox}[
    colback=usercolor!10,
    colframe=usercolor!80,
    title={\small\textbf{User Message} -- Collaborative Context with SID-grounded Items},
    fonttitle=\bfseries\color{white},
    colbacktitle=usercolor,
    boxrule=0.4pt,
    rounded corners,
    left=3pt, right=3pt, top=2pt, bottom=2pt
]
\small\ttfamily
\{\par
\hspace{1em}"role": "user",\par
\hspace{1em}"content": "\par
\vspace{0.3em}

\hspace{2em}\textbf{// Purchase History (SID + Title + Categories)}\par
\hspace{2em}The user has purchased the following items:\par
\hspace{2em}{\color{sidcolor}<|sid\_begin|><s\_a\_57>...<|sid\_end|>}, title: ``...'', categories: ``...'';\par
\hspace{2em}{\color{sidcolor}<|sid\_begin|><s\_a\_7>...<|sid\_end|>}, title: ``...'', categories: ``...'';\par
\vspace{0.3em}

\hspace{2em}\textbf{// Candidate Set (SID + Title + Categories)}\par
\hspace{2em}Please re-rank the following candidates:\par
\hspace{2em}Candidate 1: {\color{sidcolor}<|sid\_begin|>...<|sid\_end|>}, title: ``...'', categories: ``...''\par
\hspace{2em}Candidate 2: {\color{sidcolor}<|sid\_begin|>...<|sid\_end|>}, title: ``...'', categories: ``...''\par
\hspace{2em}{\color{gray}... (Candidates 3--10)}\par
\hspace{1em}"\par
\}\par
\vspace{0.3em}
\rmfamily\color{gray}\textit{$\triangleright$ Provides complete context: user history + all candidates, each with semantic ID, title, and category metadata}
\end{tcolorbox}

\vspace{0.5em}

%% ASSISTANT MESSAGE
\begin{tcolorbox}[
    colback=assistantcolor!10,
    colframe=assistantcolor!80,
    title={\small\textbf{Assistant Message} -- Structured Reasoning Trace \& Ranked Output},
    fonttitle=\bfseries\color{white},
    colbacktitle=assistantcolor,
    boxrule=0.4pt,
    rounded corners,
    left=3pt, right=3pt, top=2pt, bottom=2pt
]
\small\ttfamily
\{\par
\hspace{1em}"role": "assistant",\par
\hspace{1em}"content": "\par
\vspace{0.2em}

\vspace{0.2em}

\hspace{2em}{\color{structcolor}\{}\par
\hspace{3em}{\color{structcolor}"explanation"}: "{\color{reasoncolor}\par
\hspace{4em}\textbf{Step 1 History Summarization}: ``The user has purchased {\color{sidcolor}<|sid|>} and {\color{sidcolor}<|sid|>}...''\par
\hspace{4em}\textbf{Step 2 Category Pattern}: ``Beauty > Hair Care''\par
\hspace{4em}\textbf{Step 3 Reasoning}: ``The user prefers argan oil products...''\par
\hspace{4em}Based on patterns, Candidate 1 best matches...}",\par
\vspace{0.2em}
\hspace{3em}{\color{structcolor}"recommendations"}: {\color{structcolor}["1", "2", "3", ..., "10"]}\par
\hspace{2em}{\color{structcolor}\}}\par
\hspace{1em}"\par
\}\par
\vspace{0.3em}
\rmfamily\color{gray}\textit{$\triangleright$ Produces SID-grounded reasoning trace + ranked recommendation list in structured JSON format}
\end{tcolorbox}

\end{tcolorbox}

\vspace{1em}

The structured reasoning trace, i.e., embedded in the Assistant Message, is a key component for enabling reasoning capability for the re-ranking task with post-training, and we will discuss the corresponding generation strategy we explored in the next section.

\subsection{Reasoning Trace Generation}
\label{Reasoning Trace Generation}

To generate high-quality reasoning trace, we consider two generation strategies for the reasoning trace: \emph{targeted sampling} and \emph{rejection sampling}, depending on whether the ground truth is contained in the prompt.

\paragraph{Targeted Sampling.}
In the targeted approach, we provide the target item $\mathbf{s}_{v_{n+1}}$ and most recent $k$ historical items, then query a larger LLM to generate a reasoning trace $\tau$ explaining the target interaction:
\begin{equation}
    \tau \sim P_\theta(\cdot | \mathcal{P}_{\text{targeted}} ([\mathbf{s}_{v_1},...,\mathbf{s}_{v_k}], [\mathbf{s}_{y_1},...,\mathbf{s}_{y_c}], \mathbf{s}_{v_{n+1}})),
\end{equation}
where $\mathcal{P}_{\text{targeted}}(x, y, z)$ constructs a prompt to query the rationale for why a user who interacts with item sequence $x$ would be most interested in $z$ from candidate list $y$. 
As the ground truth is involved, the targeted approach always yields rationales for why the target might be favored by the user; however, it may hallucinate without genuine belief in the result. 

\paragraph{Rejection Sampling.}
Alternatively, in our rejection sampling strategy, we do not provide the ground truth but keep querying the LLM to predict which item $\mathbf{\hat{s}}_{y_c}$ among the pre-ranked candidates is most likely to be the user's next interest, until the prediction matches the actual target:
\begin{equation}
    \quad 
    (\tau, \mathbf{\hat{s}}_{y_c}) \sim P_\theta(\cdot \mid \mathcal{P}_{\text{rejection}}([\mathbf{s}_{v_1}, \ldots, \mathbf{s}_{v_k}], [\mathbf{s}_{y_1}, \ldots, \mathbf{s}_{y_c}]))
    \quad \text{subject to} \quad \mathbf{\hat{s}}_{y_c} = \mathbf{s}_{v_{n+1}},
\end{equation}
where $\mathcal{P}_{\text{rejection}}(x,y)$ constructs a prompt to query which item in candidate list $y$ is most likely to be the user's next interest, given history $x$.

To generate high quality reasoning trace for re-ranking, we leverage following principles to design the prompt for instructing LLM for reasoning:
\begin{itemize}
    \item {\color{principleA}\textbf{Concrete System Role and Re-ranking Task definition}}: To incentive the re-ranking capability of large LLMs, we elucidate the recommendation domain, the dedicated roles for the LLM to display, and the concrete definition of the re-ranking task.
    \item {\color{principleE}\textbf{Collaborative Context Presentation}}: To provide comprehensive decision context, we present both the user's purchase history (with structured metadata including titles and categories) and the complete candidate set simultaneously, enabling the model to perform holistic comparison rather than isolated item evaluation.
    \item {\color{principleD}\textbf{Domain Knowledge Priming}}: To leverage sequential purchase patterns inherent in e-commerce (e.g., shampoo $\rightarrow$ conditioner $\rightarrow$ styling products), we explicitly prompt the model to consider such domain-specific heuristics, enabling it to apply common-sense reasoning about product complementarity and routine-based purchasing behavior.
    \item {\color{principleB}\textbf{Critical Guidelines as Output Constraint}}: To ensure reasoning traces are grounded and verifiable, we impose explicit constraints such as requiring the model to cite items by their SIDs. This forces the model to anchor its reasoning in specific historical items rather than generating vague or hallucinated justifications, enabling direct traceability between reasoning steps and input data.
    \item {\color{principleC}\textbf{Structured Multi-Step Reasoning Format}}: To elicit progressive and interpretable reasoning, we provide an explicit step-by-step output format with examples. This hierarchical structure guides the model through: (1) broad pattern recognition, (2) identifying complementary product types, and (3) matching to specific candidates, mirroring human decision-making processes.
\end{itemize}

\vspace{0.8em}

%% MAIN PROMPT STRUCTURE
\begin{tcolorbox}[
    colback=gray!8,
    colframe=gray!60,
    title=\textbf{High-Level Prompt Structure},
    fonttitle=\bfseries,
    boxrule=0.5pt,
    rounded corners,
    breakable,
    enhanced
]

%% Principle A: System Role & Task Definition
\begin{tcolorbox}[
    colback=principleA!10,
    colframe=principleA!80,
    title={\small\textbf{P1: System Role \& Task Definition}},
    fonttitle=\bfseries\color{white},
    colbacktitle=principleA,
    boxrule=0.4pt,
    rounded corners,
    left=3pt, right=3pt, top=2pt, bottom=2pt
]
\small\ttfamily
You are an \textbf{expert at analyzing e-commerce purchase patterns} and predicting user preferences.\par\vspace{0.3em}
Given the user's purchase history and a list of candidate items, you need to predict which candidate is \textbf{MOST LIKELY} to be the user's next purchase.
\end{tcolorbox}

\vspace{0.5em}

%% Principle E: Collaborative Context
\begin{tcolorbox}[
    colback=principleE!10,
    colframe=principleE!80,
    title={\small\textbf{P2: Collaborative Context Presentation}},
    fonttitle=\bfseries\color{white},
    colbacktitle=principleE,
    boxrule=0.4pt,
    rounded corners,
    left=3pt, right=3pt, top=2pt, bottom=2pt
]
\small\ttfamily
\textbf{=== USER PURCHASE HISTORY ===}\par
- {\color{sidcolor}<|sid\_begin|>...<|sid\_end|>}; Title: [Product A]; Categories: [...]\par
- {\color{sidcolor}<|sid\_begin|>...<|sid\_end|>}; Title: [Product B]; Categories: [...]\par
{\color{gray}\textit{(Complete history with SIDs, titles, and category metadata)}}\par\vspace{0.4em}

\textbf{=== CANDIDATE ITEMS ===}\par
Candidate 1: Title: [...]; Categories: [...]\par
Candidate 2: Title: [...]; Categories: [...]\par
{\color{gray}\textit{(Full candidate set for holistic comparison)}}
\end{tcolorbox}

\vspace{0.5em}

%% Principle D: Domain Knowledge Priming
\begin{tcolorbox}[
    colback=principleD!10,
    colframe=principleD!80,
    title={\small\textbf{P3: Domain Knowledge Priming}},
    fonttitle=\bfseries\color{white},
    colbacktitle=principleD,
    boxrule=0.4pt,
    rounded corners,
    left=3pt, right=3pt, top=2pt, bottom=2pt
]
\small\ttfamily
\textbf{=== TASK ===}\par
Analyze the user's purchase history and predict which candidate they are most likely to purchase next.\par
Think step-by-step about the patterns and preferences shown in their history.\par\vspace{0.3em}
{\color{gray}\textit{Hint: Consider sequential purchase patterns (e.g., shampoo $\rightarrow$ conditioner $\rightarrow$ styling)}}
\end{tcolorbox}

\vspace{0.5em}

%% Principle B: Output Constraints
\begin{tcolorbox}[
    colback=principleB!10,
    colframe=principleB!80,
    title={\small\textbf{P4: Critical Guidelines as Output Constraints}},
    fonttitle=\bfseries\color{white},
    colbacktitle=principleB,
    boxrule=0.4pt,
    rounded corners,
    left=3pt, right=3pt, top=2pt, bottom=2pt
]
\small\ttfamily
\textbf{CRITICAL GUIDELINES:}\par
1. \textbf{CITE ITEMS BY SID}: When referring to items in purchase history, cite them directly using their SID (e.g., {\color{sidcolor}<|sid\_begin|><s\_a\_99>...<|sid\_end|>}).\par
\hspace{1em}{\color{gray}\textit{$\rightarrow$ Enables grounded, verifiable reasoning traces}}\par
2. Focus on analyzing patterns in the user's purchase history\par
3. Think about sequential purchase patterns
\end{tcolorbox}

\vspace{0.5em}

%% Principle C: Multi-Step Reasoning Format
\begin{tcolorbox}[
    colback=principleC!10,
    colframe=principleC!80,
    title={\small\textbf{P5: Structured Multi-Step Reasoning Format}},
    fonttitle=\bfseries\color{white},
    colbacktitle=principleC,
    boxrule=0.4pt,
    rounded corners,
    left=3pt, right=3pt, top=2pt, bottom=2pt
]
\small\ttfamily
\textbf{=== EXAMPLE OUTPUT FORMAT ===}\par
\textbf{Step 1}: ``Looking at the purchase history, {\color{sidcolor}<|sid|>} and {\color{sidcolor}<|sid|>} are both hair care products...''\par
\hspace{1em}{\color{gray}\textit{$\rightarrow$ Broad pattern recognition}}\par\vspace{0.2em}
\textbf{Step 2}: ``The recent purchase of {\color{sidcolor}<|sid|>} suggests the user is looking for...''\par
\hspace{1em}{\color{gray}\textit{$\rightarrow$ Identify complementary product types}}\par\vspace{0.2em}
\textbf{Step 3}: ``Based on this pattern, Candidate X would be the natural next purchase...''\par
\hspace{1em}{\color{gray}\textit{$\rightarrow$ Match to specific candidate}}\par\vspace{0.2em}
\textbf{Prediction}: Re-ranked List
\end{tcolorbox}

\end{tcolorbox}

\section{Reasoning Enablement for Re-Ranking}\label{sec:stage_3}

\subsection{Problem Setup and Prompt Interface}
We study the problem of \emph{reasoning-enabled re-ranking} over a fixed candidate set.
Given a user purchase history and a pre-ranked list of candidate items produced by a retriever,
the goal is to re-rank the candidates such that the ground-truth next item is promoted.

Each training or inference instance is formatted as a chat prompt consisting of three roles:
\texttt{system}, \texttt{user}, and \texttt{assistant}. We follow the same role semantics
as defined in Section~\ref{sec:stage_2}.
During supervised fine-tuning (SFT), the model is trained to generate the full assistant message.
During reinforcement learning (RL), the policy conditions only on the system and user messages,
and the assistant output is treated as the action.

\subsection{Supervised Fine-tuning with Reasoning Traces}
Recommender models often struggle to generate effective CoT reasoning from noisy and lengthy real-world user behavior sequences~\citep{liu2025onerec}. The generated reasoning traces are used to supervise a base LLM to acquire reasoning capability for re-ranking. We fine-tune the model to generate both the reasoning trace $\tau = [r_1, \ldots, r_M]$ and the ranked output $\mathbf{o} = [o_1, \ldots, o_T]$.

To preserve ranking performance while enabling reasoning, we decouple the losses for reasoning and ranking tokens. Specifically, we apply the language modeling loss \emph{only to the assistant message}, with separate weights for reasoning and ranking segments:
\begin{equation}
\mathcal{L}_{\text{SFT}} =
- \lambda_r \sum_{i=1}^{M}
\log P(r_i \mid \mathcal{P}, r_{<i})
- \lambda_o \sum_{j=1}^{T}
\log P(o_j \mid \mathcal{P}, \tau, o_{<j}),
\end{equation}
where $\lambda_r < \lambda_o$ balances reasoning fluency and ranking accuracy. This training procedure teaches the model to generate SID-grounded, coherent reasoning traces that connect user history with candidate comparisons, while maintaining strong re-ranking behavior.

\subsection{Reinforcement Learning}

While SFT enables coherent reasoning, it does not directly optimize the re-ranking objective.
Building upon the distilled reasoning capabilities, we apply reinforcement learning (RL) to further refine both the reasoning process and the final ranking quality. Given a prompt $\mathcal{P}(\mathcal{H}, \mathcal{D})$, the policy $\pi_\theta$ generates an output $o = (\tau, \mathbf{o})$. 
During RL, the policy conditions only on the system and user messages; the assistant output is treated as the action. 
This stage aims to lift the rank of target item in the pre-ranked list using a novel reward mechanism tailored for generative recommendation.

\subsubsection{Ranking Reward}
We define a ranking reward that measures how much the target item is promoted by re-ranking:
\begin{equation}
R_{\text{rank}} = \frac{ r^{\mathcal{D}}_{\mathbf{s}_{v_{n+1}}} - r^{\mathbf{o}}_{\mathbf{s}_{v_{n+1}}}}{|\mathcal{D}|},
\end{equation}
where $r^{\mathcal{D}}_{\mathbf{s}_{v_{n+1}}}$ and $r^{\mathbf{o}}_{\mathbf{s}_{v_{n+1}}}$ denote the ranks of the target item in the pre-ranked and re-ranked lists, respectively.

\subsubsection{Conditional Format Reward}
To ensure parseable outputs, we introduce a format reward $R_{\text{fmt}} = \Omega(o)$,
which checks whether both the reasoning trace $\tau$ and ranking output $\mathbf{o}$ can be reliably extracted. 
The details of the parsing function is illustrated in Appendix~\ref{sec:appendix method}. However, naively combining $R_{\text{rank}}$ and $R_{\text{fmt}}$
may lead to reward hacking, where the model preserves the original ranking to obtain format reward.
We therefore apply the format reward conditionally:
\begin{equation}
R =
\begin{cases}
R_{\text{rank}} + \alpha R_{\text{fmt}},
& \text{if } R_{\text{rank}} > 0
\ \text{or}\ r^{\mathcal{D}}_{\mathbf{s}_{v_{n+1}}}=1, \\
R_{\text{rank}}, & \text{otherwise}.
\end{cases}
\end{equation}

Here $\alpha$ is a weight hyperparameter of the format reward.

\subsubsection{Training via DAPO}

We optimize the policy using Decoupled Clip and Dynamic sAmpling Policy Optimization (DAPO) algorithm~\citep{yu2025dapo}, a recently state-of-the-art RL algorithm developed upon Grouped Policy Optimization
(GRPO)~\citep{shao2024deepseekmath}. 
It can effectively resolve the entropy collapse phenomenon and rollout length bias identified in the GRPO training process. For each prompt, we sample a group of $G$ outputs
$\{o_i\}_{i=1}^G$
and optimize:
\begin{equation}
\mathcal{J}_{\text{DAPO}}(\theta) = \mathbb{E}_{(q,D)\sim\mathcal{D},\{o_i\}_{i=1}^G \sim \pi_{\theta_{\text{old}}}(\cdot|q)} \left[ \frac{1}{\sum_{i=1}^G |o_i|} \sum_{i=1}^G \sum_{t=1}^{|o_i|} \min\left(r_{i,t}(\theta) \hat{A}_{i,t}, \text{clip}(r_{i,t}(\theta), 1 - \varepsilon_{\text{low}}, 1 + \varepsilon_{\text{high}}) \hat{A}_{i,t}\right) \right]
\label{eq:8}
\end{equation}
s.t. $0 < |\{o_i | \text{is\_equivalent}(a, o_i)\}| < G$,
where
\begin{equation}
r_{i,t}(\theta) = \frac{\pi_\theta(o_{i,t} | q, o_{i,<t})}{\pi_{\theta_{\text{old}}}(o_{i,t} | q, o_{i,<t})}, \quad \hat{A}_{i,t} = \frac{R_i - \text{mean}(\{R_i\}_{i=1}^G)}{\text{std}(\{R_i\}_{i=1}^G)}.
\label{eq:9}
\end{equation}
$\pi_\theta$ and $\pi_{\theta_{\text{old}}}$ denote the current and previous policies, respectively. 
The pair $(q,D)\sim\mathcal{D}$ is sampled from the training distribution. The advantage estimate $\hat{A}_{i,t}$ is computed for each output sequence $o_i$ and normalized within the group $G$. 
Adhering to the Clip-Higher strategy, DAPO decouples the lower and higher clipping range as $\varepsilon_{\text{low}}$ and $\varepsilon_{\text{high}})$. 
Additionally, to avoid zero policy gradients (advantages) and improve sample efficiency, it oversamples and filters out prompts with the accuracy equal to 1 and 0.
\section{Experiments}\label{sec:exp}
\label{experiments}
In this section, we conduct comprehensive experiments to answer the following research questions:

\begin{itemize}
    \item{\textbf{Q1}}: What techniques are essential for generating semantic IDs with high uniqueness, and how do different codebook configurations affect downstream recommendation performance?
    \item{\textbf{Q2}}: Does the proposed tokenization enhancement (Section~\ref{sec:contrastive-loss} and~\ref{sec:codebook-techniques}) and mid-training strategy improve recommendation quality compared to existing state-of-the-art LLM-based approaches?
    \item{\textbf{Q3}}: How do different reasoning trace generation strategies (targeted vs. rejection sampling) and training paradigms (SFT vs. RL) affect re-ranking performance, and is reinforcement learning necessary for translating reasoning capabilities into re-ranking improvements?
\end{itemize}

\subsection{Experiment Setup}

\subsubsection{\textbf{Datasets}}
\label{Datasets}

We evaluate our models on two Amazon Review datasets (Beauty and Sports) derived from the large-scale Amazon product review corpus curated by \cite{mcauley2015image}, which contains user–item interactions with rich side information such as ratings, timestamps, and product metadata. This corpus has been widely adopted as a benchmark for recommendation research due to its scale, diversity of product domains, and realistic user behavior logs. The statistics of the datasets are presented in Table~\ref{tab:dataset_statistics}. All datasets are preprocessed following standard protocols~\citep{rajput2023recommender}: users and items with fewer than $5$ interactions are filtered out\footnote{\url{https://cseweb.ucsd.edu/~jmcauley/datasets/amazon/links.html}}
, interactions are sorted chronologically for each user, and a leave-one-out strategy is employed to construct validation and test sets.

\begin{table}[t]
\centering
% \small
\caption{Statistics of the Amazon Review datasets used for sequential recommendation.}
\begin{tabular}{lcccc}
\toprule
\textbf{Dataset} & \textbf{\#Users} & \textbf{\#Items} & \textbf{Avg. Seq. Len.} \\
\midrule
Beauty & $22,363$ & $12,101$ & $8.87$ \\
Sports & $35,598$ & $18,357$ & $8.32$ \\
% Toys   & $19,412$ & $11,924$ & $8.63$ \\
\bottomrule
\end{tabular}
\label{tab:dataset_statistics}
\end{table}

\subsubsection{\textbf{Metrics}}
\label{sec:metrics}

We evaluate sequential recommendation performance using standard metrics: \emph{Recall@K} and \emph{Normalized Discounted Cumulative Gain at K (NDCG@K)}.

To evaluate the quality of the learned SIDs, we measure the \emph{Collision/Uniqueness Rate}, which quantifies how often multiple items are assigned the same SID.

Detailed introduction to the metrics are presented in Appendix~\ref{sec:appendix metrics}.

\subsection{Tokenizer Selection}

\subsubsection{High SID Uniqueness}
\begin{table}[htbp]
\centering
\caption{Ablation study on techniques for improving codebook uniformity in RQ-VAE. 
We report unique Uniqueness rate (\%) on the Amazon Beauty dataset. 
Configurations achieving $\geq$99\% are \textbf{bolded}.}
\label{tab:ablation}
\begin{tabular}{ccccc|c}
\toprule
\textbf{Diversity} & \textbf{Dead Code} & \textbf{EMA} & \textbf{Random} & \textbf{Contrastive} & \textbf{Uniqueness Rate} \\
\textbf{Loss} & \textbf{Reset} & \textbf{Update} & \textbf{Last Level} & \textbf{Loss} & \textbf{(\%)} \\
\midrule
\multicolumn{6}{c}{\textit{Baseline Configurations (EMA Enabled)}} \\
\midrule
 & \cmark & \cmark &  &  & 98.77 \\
 & \cmark & \cmark &  & \cmark & 98.28 \\
 & \cmark & \cmark & \cmark &  & \textbf{99.98}  \\
 & \cmark & \cmark & \cmark & \cmark & \textbf{99.98}  \\
\cmark & \cmark & \cmark &  &  & 98.92 \\
\cmark & \cmark & \cmark &  & \cmark & 97.89 \\
\cmark & \cmark & \cmark & \cmark &  & \textbf{99.98}  \\
\cmark & \cmark & \cmark & \cmark & \cmark & \textbf{99.95}  \\
\midrule
\multicolumn{6}{c}{\textit{Without Dead Code Reset (EMA Enabled)}} \\
\midrule
 &  & \cmark &  &  & 93.74 \\
 &  & \cmark &  & \cmark & 95.92 \\
 &  & \cmark & \cmark &  & \textbf{99.67}  \\
 &  & \cmark & \cmark & \cmark & \textbf{99.84}  \\
\cmark &  & \cmark &  &  & 94.51 \\
\cmark &  & \cmark &  & \cmark & 95.92 \\
\cmark &  & \cmark & \cmark &  & \textbf{99.78}  \\
\cmark &  & \cmark & \cmark & \cmark & \textbf{99.86}  \\
\midrule
\multicolumn{6}{c}{\textit{Without EMA Update (Codebook Collapse)}} \\
\midrule
 & \cmark &  &  &  & 21.39 \\
 & \cmark &  &  & \cmark & 6.35 \\
 & \cmark &  & \cmark &  & 66.26 \\
 & \cmark &  & \cmark & \cmark & 32.72 \\
 &  &  &  &  & 21.76 \\
 &  &  &  & \cmark & 3.91 \\
 &  &  & \cmark &  & 60.71 \\
 &  &  & \cmark & \cmark & 51.43 \\
\cmark & \cmark &  &  &  & 21.35 \\
\cmark & \cmark &  &  & \cmark & 3.16 \\
\cmark & \cmark &  & \cmark &  & 59.17 \\
\cmark & \cmark &  & \cmark & \cmark & 45.62 \\
\cmark &  &  &  &  & 19.64 \\
\cmark &  &  &  & \cmark & 4.54 \\
\cmark &  &  & \cmark &  & 68.39 \\
\cmark &  &  & \cmark & \cmark & 40.85 \\
\bottomrule
\end{tabular}
\end{table}

We conduct an ablation study to analyze the impact of each technique on SID uniqueness, whose results are presented in Table~\ref{tab:ablation}. 
Our findings reveal two categories of techniques based on their effectiveness:

\begin{itemize}
    \item \textbf{Essential Techniques:}
    \begin{itemize}
        \item \textbf{EMA Update} is critical for preventing codebook collapse. 
              Enabling EMA improves the average SID rate from 31.32\% to 97.50\% (+66.2\%), 
              making it the most impactful technique.
        \item \textbf{Random Last Level} significantly boosts uniqueness by randomly assigning 
              the final quantization level during inference. This technique improves the 
              average SID rate from 54.06\% to 82.76\% (+28.7\%), and all configurations 
              achieving $\geq$99\% uniqueness use this technique.
    \end{itemize}
    
    \item \textbf{Optional Techniques (marginal or negative effect):}
    \begin{itemize}
        \item \textbf{Diversity Loss} ($\lambda$=0.1) has minimal impact (+0.6\%), 
              suggesting that EMA updates already provide sufficient codebook utilization.
        \item \textbf{Dead Code Reset} ($\tau$=2) shows a slight negative effect ($-$4.0\%), 
              indicating that resetting unused codes may disrupt learned representations.
        \item \textbf{Contrastive Loss} unexpectedly hurts uniqueness ($-$13.2\%), 
              likely because it encourages similar items to share codes, 
              increasing collisions.
    \end{itemize}
\end{itemize}

In summary, achieving $\geq$99\% unique SID rate requires both EMA Update and Random Last Level, 
while the other techniques provide marginal or even negative contributions.

\subsubsection{Best-Performing SID}
While achieving a high unique SID rate ($\geq$99\%) is necessary to ensure each item receives a distinct identifier, it does not guarantee that the resulting semantic IDs are \emph{semantically meaningful} for downstream recommendation tasks. 
To evaluate the quality of different codebook configurations, we train a TIGER-style~\citep{rajput2023recommender} sequential recommendation model on the Amazon Beauty dataset. 
The model takes a user's purchase history---represented as a sequence of semantic IDs---and autoregressively predicts the next item's semantic ID using beam search decoding. 
We then convert the predicted SIDs back to raw item IDs and compute standard ranking metrics (Recall@$K$ and NDCG@$K$) against the ground-truth next item. Table~\ref{tab:ablation_cookbook} reports results for all eight configurations that achieved $\geq$99\% unique SID rate. 
We observe that \textbf{contrastive loss consistently improves recommendation performance} across all settings, with the best configuration (no diversity loss, dead code reset enabled, EMA update, random last level, and contrastive loss) achieving Recall@10 of 0.0367 and NDCG@10 of 0.0190. 
In the other datasets, we follow this best-performing config.

\begin{table}[htbp]
\centering
\caption{Ablation study on codebook balancing techniques for RQ-VAE on Amazon Beauty dataset. All configurations use EMA updates and random last level assignment, which are required to achieve $\geq$99\% unique SID rate. Metrics are evaluated using a TIGER-style sequential recommendation model trained for 50 epochs.}
\label{tab:ablation_cookbook}
\begin{tabular}{ccccc|cccc}
\toprule
\textbf{Diversity} & \textbf{Dead Code} & \textbf{EMA} & \textbf{Random} & \textbf{Contrastive} & \textbf{Recall} & \textbf{Recall} & \textbf{NDCG} & \textbf{NDCG} \\
\textbf{Loss} & \textbf{Reset} & \textbf{Update} & \textbf{Last Level} & \textbf{Loss} & \textbf{@5} & \textbf{@10} & \textbf{@5} & \textbf{@10} \\
\midrule
            & \checkmark & \checkmark & \checkmark &            & 0.0197 & 0.0314 & 0.0126 & 0.0164 \\
            & \checkmark & \checkmark & \checkmark & \checkmark & \textbf{0.0227} & \textbf{0.0367} & \textbf{0.0145} & \textbf{0.0190} \\
            &            & \checkmark & \checkmark &            & 0.0187 & 0.0281 & 0.0116 & 0.0146 \\
            &            & \checkmark & \checkmark & \checkmark & 0.0209 & 0.0343 & 0.0128 & 0.0171 \\
\checkmark  & \checkmark & \checkmark & \checkmark &            & 0.0204 & 0.0314 & 0.0131 & 0.0167 \\
\checkmark  & \checkmark & \checkmark & \checkmark & \checkmark & 0.0226 & 0.0359 & 0.0138 & 0.0181 \\
\checkmark  &            & \checkmark & \checkmark &            & 0.0198 & 0.0308 & 0.0123 & 0.0158 \\
\checkmark  &            & \checkmark & \checkmark & \checkmark & 0.0212 & 0.0336 & 0.0132 & 0.0172 \\
\bottomrule
\end{tabular}
\end{table}

\subsection{Results of the Tokenized Mid-Training}

In Table~\ref{tab:mid-training} we demonstrate the performance improvement achieved by our tokenization enhancement and the mid-training compared to the existing state-of-the-art, i.e., OneRec-Think~(ORT)~\citep{liu2025onerec}, on the Amazon Beauty dataset. 
The ``Base'' and ``Base+IA'' are defined the same as in the OneRec-Think~(ORT)~\citep{liu2025onerec} paper, i.e., ``Base'' denotes the model tuned by the raw itemic token sequence while the ``Base+IA'' denotes the model enhanced with Itemic Alignment. 
These are two different training approach to empower the recommendation capability on LLM before the reasoning-based post-training. We observe that: (1) our tokenization enhancement alone yields consistent improvements over the ORT Base'' model across all metrics, with relative gains of 6.7\% on recall@5, 6.3\% on recall@10, 4.5\% on ndcg@5, and 4.2\% on ndcg@10, demonstrating that our semantic tokenization strategy better captures item representations; 
(2) when combined with mid-training, both our single-task and multi-task learning approaches substantially outperform the ORT Base+IA'' baseline on recall@5, recall@10, and ndcg@5, with the most notable improvement being 18.7\% on ndcg@5 for single-task learning; 
(3) multi-task learning, despite training on additional tasks beyond sequential preference prediction, achieves on-par or superior results, notably a 5.3\% improvement on recall@10, highlighting the success of the multi-task training framework in balancing diverse objectives without sacrificing performance on the primary recommendation task.

\begin{table}[htbp]
  \centering
  \caption{Performance comparison with the OneRec-Think~(ORT)~\citep{liu2025onerec} on the Amazon Beauty dataset. Our proposed tokenization enhancement and the corresponding mid-training (Section~\ref{sec:stage_1}) achieve significant improvement over the ORT's results provided by their official GitHub implementation~(https://github.com/wangshy31/OneRec-Think). The result of ``Base'' and ``Base+IA'' are cited from the Table 2 of the ORT paper.}
  \label{tab:results}
  \begin{tabular}{lcccc}
  \toprule
  Methods & recall@5 & recall@10 & ndcg@5 & ndcg@10 \\
  \midrule
  Base (ORT~\citep{liu2025onerec}) & 0.0460 & 0.0654 & 0.0314 & 0.0377 \\
  tokenization enhancement~(Ours) & \textbf{0.0491} & \textbf{0.0695} & \textbf{0.0328} & \textbf{0.0393} \\
  \midrule
  Base+IA (ORT~\citep{liu2025onerec}) & 0.0532 & 0.0735 & 0.0342 & \textbf{0.0471} \\
  Mid-Training w/ Single-task~(Ours) & \textbf{0.0564} & 0.0744 & \textbf{0.0406} & 0.0467 \\
  Mid-Training w/ Multi-task~(Ours) & 0.0561 & \textbf{0.0774} & 0.0396 & 0.0467 \\
  \bottomrule
  \label{tab:mid-training}
  \end{tabular}
\end{table}

\subsection{Results of Re-Ranking}

\begin{table}[htbp]
  \centering
  \caption{Re-ranking performance comparison with retriever on the Amazon Beauty dataset. Our post-training achieve significant improvement over pre-ranked results. ``MTL'' denotes mid-training with multi-task. ``RL-zeroshot'' means RL directly from the base model without SFT. ``targeted'' and ``rejection'' correspond to the two reasoning trace generation strategies used to SFT and serve as the reference policies in RL stage. ``KP'' denotes domain knowledge priming in reasoning generation.}
  \label{tab:post_training_results_beauty}
  \begin{tabular}{lccccc}
  \toprule
  Methods & recall@1 & recall@5 & recall@9 & ndcg@5 & ndcg@10 \\
  \midrule
  Pre-rank (MTL) & 0.2892 & 0.7227 & 0.9534 & 0.5101 & 0.5997 \\
  \midrule
  SFT-targeted-KP & 0.2006 & 0.7250 & 0.9545 & 0.4759 & 0.5647 \\
  SFT-targeted-noKP & 0.2761 & 0.7272 & 0.9500 & 0.5087 & 0.5964 \\
  SFT-rejection-KP & 0.2784 & 0.7091 & 0.9483 & 0.4970 & 0.5907 \\
  SFT-rejection-noKP & 0.2682 & 0.7216 & 0.9574 & 0.5003 & 0.5904 \\
  \midrule
  RL-zeroshot & 0.2801 & 0.7074 & 0.9574 & 0.4982 & 0.5926 \\
  RL-targeted-KP & 0.2898 & 0.7221 & 0.9534 & 0.5101 & 0.5998 \\
  RL-targeted-noKP & 0.2932 & 0.7318 & 0.9534 & 0.5172 & 0.6038 \\
  RL-rejection-KP & \textbf{0.2977} & \textbf{0.7460} & 0.9563 & \textbf{0.5234} & \textbf{0.6050} \\
  RL-rejection-noKP & 0.2915 & 0.7330 & \textbf{0.9580} & 0.5172 & 0.6036 \\
  \bottomrule
  \end{tabular}
\end{table}

\begin{table}[htbp]
  \centering
  \caption{Re-ranking performance comparison with retriever on the Amazon Sports dataset. ``STL'' denotes mid-training with single-task.}
  \begin{tabular}{lccccc}
  \toprule
  Methods & recall@1 & recall@5 & recall@9 & ndcg@5 & ndcg@10 \\
  \midrule
  Pre-rank (STL) & 0.2422 & 0.7077 & 0.9583 & 0.4796 & 0.5742 \\
  \midrule
  SFT-targeted-KP & 0.1790 & 0.7025 & 0.9557 & 0.4491 & 0.5454 \\
  SFT-targeted-noKP & 0.2279 & 0.7103 & \textbf{0.9596} & 0.4750 & 0.5688 \\
  SFT-rejection-KP & 0.2363 & 0.7090 & 0.9525 & 0.4723 & 0.5664 \\
  SFT-rejection-noKP & 0.2233 & 0.6608 & 0.9427 & 0.4437 & 0.5532 \\
  \midrule
  RL-zeroshot & \textbf{0.2461} & 0.7012 & 0.9447 & 0.4772 & 0.5732 \\
  RL-targeted-KP & 0.2428 & \textbf{0.7083} & 0.9583 & \textbf{0.4802} & \textbf{0.5745} \\
  RL-targeted-noKP & 0.2383 & \textbf{0.7083} & 0.9583 & 0.4786 & 0.5730 \\
  RL-rejection-KP & 0.2415 & 0.7077 & 0.9583 & 0.4791 & 0.5737 \\
  RL-rejection-noKP & 0.2370 & 0.7044 & 0.9583 & 0.4743 & 0.5701 \\
  \bottomrule
  \label{tab:post_training_results_sports}
  \end{tabular}
\end{table}

The model obtained from mid-training is used as a \emph{retriever} to produce a top-$K$ pre-ranked candidate list for each prompt, which also serves as the baseline. 
The goal of the proposed GR2 framework is to improve recommendation quality by \emph{re-ranking} these candidates through reasoning-aware post-training. 
Table~\ref{tab:post_training_results_beauty} and Table~\ref{tab:post_training_results_sports} report the re-ranking results on the Amazon Beauty and Sports datasets, respectively, with $K=10$.

For each dataset, the pre-rank baseline is selected based on recall@10 performance. 
Specifically, mid-training with multi-task learning achieves the best recall@10 on the Beauty dataset and is therefore used to generate the candidate lists for re-ranking, while the single-task mid-trained model is used for the Sports dataset. 
Both the retriever and the re-ranker are initialized from Qwen3-8B. We make the following observations:
(1) Reasoning trace quality plays a critical role in re-ranking performance: on the Beauty dataset, rejection sampling with hierarchical item information (RL-untargeted-KP) consistently outperforms vanilla reasoning traces (RL-zeroshot) under the same RL setting, yielding relative improvements of 3.00\% in recall@1, 3.30\% in recall@5, and 2.60\% in ndcg@5. 
This demonstrates that structured and informative reasoning traces provide a stronger learning signal for ranking refinement; 
(2) RL further enhances reasoning-activated models: when applied on top of reasoning-aware SFT, on the sports dataset, RL-targeted-KP improves over the pre-rank baseline by 0.24\% recall@1 and 0.13\% ndcg@5, indicating that policy optimization can refine the reasoning process to better align with ranking objectives; 
(3) Improving re-ranking via reasoning is non-trivial and requires RL: although SFT learns high-quality and coherent reasoning traces, it does not consistently translate into improved re-ranking performance, and in some cases even degrades recall@1. 
This highlights the mismatch between reasoning quality and ranking optimality, and underscores the necessity of RL to explicitly optimize the reasoning action space toward ranking-aware rewards.
\section{Related Works}\label{sec:related}
\subsection{Generative LLM RecSys}
Recent work has increasingly reformulated recommendation as a sequence generation problem using large language models. 
TIGER~\citep{rajput2023recommender} pioneered this direction by introducing Semantic IDs (SIDs), which discretize item representations via RQ-VAE and enable autoregressive recommendation without large embedding tables. 
Building on the generative formulation, OneRec~\citep{zhou2025onerec} proposed a unified text-to-text framework that casts multiple recommendation tasks into a single LLM-based generation paradigm. 
To further enhance model capability, OneRec-Think~\citep{liu2025onerec} incorporated explicit reasoning traces, showing that chain-of-thought-style intermediate reasoning can improve recommendation accuracy and robustness. 
Complementarily, OpenOneRec~\citep{zhou2025openonerec} focused on reproducibility and extensibility by providing a standardized research framework for generative recommenders. 
PLUM~\citep{he2025plum} shares a similar idea to OneRec-Think in using continued pre-training to align pre-trained LLMs with recommendation domains and further introduces a task-specific fine-tuning objective for generative retrieval. 
Together, these works characterize the evolution of generative LLM-based recommenders, from representation tokenization to reasoning- and planning-aware generation.

In contrast, our GR2 inherits key design elements such as semantic IDs and item alignment training, while placing special emphasis on (1) tokenizers that achieve higher uniqueness in semantic ID representations, (2) high-quality reasoning trace generation via carefully designed prompts and rejection sampling, and (3) tailored DAPO-based optimization and reward functions specifically designed for the re-ranking problem.

\subsection{Reasoning LLM}
Chain-of-thought (CoT) prompting has emerged as a foundational technique to elicit multi-step reasoning from LLMs by generating intermediate steps before final answers~\citep{wei2022chain}. Subsequent work has formalized and extended reasoning structures beyond CoT~\citep{xia2025beyond}, including strategic generation of reasoning texts to stabilize performance~\citep{wang2024strategic}. 
RL approaches such as GRPO and its variants have been shown to further enhance the faithfulness and coherence of LLM reasoning by optimizing verifiable reward signals during training~\citep{shao2024deepseekmath, yu2025dapo, zheng2025group, chen2025minimax}. These algorithms provide a broader context for our optimization strategy designed for reasoning-guided re-ranking.
LLM-based document re-ranking in search domain is the study closest our work. For instance, ReaRank explicitly reasons before re-ranking passage lists via RL, achieving improved relevance and interpretability~\citep{zhang2025rearank}. Rank-R1 enhances LLM-based rerankers through RL-based reasoning optimization on document ranking benchmarks~\citep{feng2025iranker}. Similarly, ReasonRank and MM-R5 explore reasoning-augmented reranking in multi-view and multi-modal settings~\citep{xu2025mm}, highlighting the promise of reasoning-aware ranking agents. 
$R^4ec$~\citep{gu2025r} employs LLM reasoning capability to iteratively reflection and refine recommendation.
LLMs are used to enhance re-ranking accuracy and interpretability by applying advanced reasoning and a bootstrapping mechanism that randomizes candidate lists, which has been shown to mitigate position bias and promote fairness~\citep{wang2025llm}.
Our GR2 differs in its focus on semantic IDs and tailored reward functions for recommendation re-ranking, leveraging structured reasoning traces and rejection sampling for higher-quality supervision.
\section{Conclusion}\label{sec:con}
We propose Generative Reasoning Re-ranker (GR2), a novel framework that elevates the re-ranking stage in recommendation systems by fully leveraging the reasoning capabilities of large language models. 
GR2 introduces a three-stage pipeline: (1) mid-training on highly unique semantic IDs to bridge item semantics and world knowledge, (2) generation and supervised fine-tuning on high-quality reasoning traces to impart foundational reasoning skills, and (3) reinforcement learning with a custom reward function tailored for re-ranking, ensuring robust and scalable supervision. 
Our extensive experiments on real-world datasets demonstrate that GR2 consistently surpasses strong baselines in both recall and ranking metrics, with ablation studies highlighting the importance of advanced reasoning and RL objectives. 
By integrating semantic representations, structured reasoning, and reward-driven optimization, GR2 sets a new benchmark for interpretable and effective re-ranking in large-scale recommendation systems, paving the way for future research in reasoning-aware recommendation models.

\bibliographystyle{assets/plainnat}
\bibliography{paper}

\clearpage
\newpage
\beginappendix
\section{Semantic ID (SID)}
\label{sec:appendix sid}

Given the textual feature $x$ of an item, a textual encoder produces a continuous embedding:
\begin{equation}
\mathbf{h} = f_{\text{enc}}(x), 
\quad \mathbf{h} \in \mathbb{R}^d.
\label{eq:encoder}
\end{equation}

Residual quantization is performed sequentially using $K$ codebooks, where $K$ denotes the number of quantization levels. 
For each level $k \in \{1,\ldots,K\}$, we define a codebook
$\mathcal{E}^{(k)} = \{\mathbf e^{(k)}_j\}_{j=1}^{C_k}$, where $C_k$ is the cardinality of the $k$-th codebook.
The residual is initialized as
\begin{equation}
\mathbf{r}_1 = \mathbf{h}.
\label{eq:residual-init}
\end{equation}

At the $k$-th level, nearest-neighbor quantization is applied:
\begin{align}
z_k &= \arg\min_{j \in \{1,\ldots,C_k\}} 
\left\|\mathbf{r}_k - \mathbf{e}^{(k)}_j\right\|_2^2,
\label{eq:nearest-neighbor} \\
\mathbf{q}_k &= \mathbf{e}^{(k)}_{z_k},
\label{eq:codebook-select} \\
\mathbf{r}_{k+1} &= \mathbf{r}_k - \mathbf{q}_k,
\qquad k=1,\ldots,K.
\label{eq:residual-update}
\end{align}

The reconstructed representation is obtained by summing all quantized vectors:
\begin{equation}
\hat{\mathbf{h}} = \sum_{k=1}^K \mathbf{q}_k.
\label{eq:reconstruction}
\end{equation}

The sequence of discrete indices
\begin{equation}
\boxed{
\text{Tokenizer}(x) = (z_1, z_2, \ldots, z_K),
\quad z_k \in \{1,\ldots,C_k\}
}
\end{equation}
can be regarded as the semantic ID (SID) of item $x$.

For a single sample $x$, the training objective is
\begin{align}
\mathcal L(x)
&= \mathcal L_{\text{rec}}
+ \beta \sum_{k=1}^K \mathcal L_{\text{cb}}^{(k)}
+ \gamma \sum_{k=1}^K \mathcal L_{\text{com}}^{(k)},
\label{eq:total-loss}\\
\mathcal L_{\text{rec}}
&= \left\| \mathbf{h} - \hat{\mathbf{h}} \right\|_2^2, \\
\mathcal L_{\text{cb}}^{(k)}
&= \left\| \text{sg}[\mathbf{r}_k] - \mathbf{q}_k \right\|_2^2, \\
\mathcal L_{\text{com}}^{(k)}
&= \left\| \mathbf{r}_k - \text{sg}[\mathbf{q}_k] \right\|_2^2.
\end{align}
Here, $\text{sg}[\cdot]$ denotes the stop-gradient operator.

Based on SIDs, TIGER~\citep{rajput2023recommender} reformulates recommendation as an autoregressive generation problem over discrete item representations. Given a user’s interaction history $\mathcal{H}$, a Transformer model sequentially predicts the next item’s Semantic ID $(z_1,\ldots,z_K)$:
\begin{equation}
p\big((z_1,\ldots,z_K)\mid \mathcal{H}\big) = \prod_{k=1}^K p\big(z_k \mid \mathcal{H}, z_{<k}\big).
\end{equation}

The training algorithm for the RQ-VAE is detailed in Algorithm~\ref{alg:rqvae-training}.

% Algorithm
\begin{algorithm}[t]
\caption{Vanilla RQ-VAE Training}
\raggedright
\textbf{Input:} Training set $\mathcal{X}$, codebooks $\{\mathbf{e}^{(k)}_j\}$, encoder $f_{\text{enc}}$, hyperparameters $\beta, \gamma$ \\
\textbf{Output:} Updated codebooks $\{\mathbf{e}^{(k)}_j\}$
\begin{algorithmic}[1]
\For{each sample $x \in \mathcal{X}$}
    \State Compute embedding $\mathbf{h}$ and initialize residual $\mathbf{r}_1$ using Eqs.~\eqref{eq:encoder} and~\eqref{eq:residual-init}
    \For{$k = 1$ to $K$}
        \State Generate $k$-th level SID $z_k$ using Eqs~\eqref{eq:nearest-neighbor},~\eqref{eq:codebook-select}, and~\eqref{eq:residual-update}
    \EndFor
    \State Update codebooks $\{\mathbf{e}^{(k)}_j\}$ to minimize the objective in Eq.~\eqref{eq:total-loss}
\EndFor
\end{algorithmic}
\label{alg:rqvae-training}
\end{algorithm}

At inference time, the generated Semantic ID is mapped back to candidate items via the tokenizer’s inverse mapping, enabling direct item retrieval.

\section{Examples of Item Alignment Tasks}
\label{sec:item alignment examples}

Here we present representative examples of \textbf{Item Alignment} tasks. OneRec-Think~\citep{liu2025onerec} introduces four such tasks: \textbf{(1) Interleaved User Persona Grounding, (2) Sequential Preference Modeling, (3) Itemic Dense Captioning, and (4) General Language Modeling}. An example of Itemic Dense Captioning is provided in Section~\ref{sec:item alignment}. Since the General Language Modeling task is trained on general-purpose text corpora, we omit its examples for brevity.

The following prompt illustrates one of the item alignment training tasks, namely \textbf{Itemic Dense Captioning}.

\begin{tcolorbox}[
    colback=gray!10,
    colframe=gray!60,
    title=Prompt Example for Itemic Dense Captioning,
    fonttitle=\bfseries,
    boxrule=0.5pt,
    rounded corners,
    breakable,
    enhanced
]
\small\ttfamily

\textless|im\_start|\textgreater system\\
You are an expert content analyst. Given an itemic token, generate a concise and accurate textual description of its content.\\
\textless|im\_end|\textgreater\\[0.6em]
\textless|im\_start|\textgreater user\\
Provide a description for the itemic token:\\
{\color{blue}
\textless|sid\_begin|\textgreater
\textless s\_a\_97\textgreater
\textless s\_b\_168\textgreater
\textless s\_c\_137\textgreater
\textless s\_d\_135\textgreater
\textless|sid\_end|\textgreater
}\\
\textless|im\_end|\textgreater\\[0.6em]
\textless|im\_start|\textgreater assistant\\
Title: Stella McCartney Stella\\
Description: STELLA For Women By STELLA MCCARTNEY 1.7 oz EDP Spray\\
Categories: Beauty $>$ Fragrance $>$ Women's $>$ Eau de Parfum\\
\textless|im\_end|\textgreater

\end{tcolorbox}

We next present examples of the Interleaved User Persona Grounding and Sequential Preference Modeling tasks.

\begin{tcolorbox}[
    colback=gray!10,
    colframe=gray!60,
    title=Prompt Example for Interleaved User Persona Grounding,
    fonttitle=\bfseries,
    boxrule=0.5pt,
    rounded corners,
    breakable,
    enhanced
]
\small\ttfamily

\textbf{User ID}: {\color{blue}A6F8KH0J1AVYA}\\[0.6em]
\textbf{\# User Profile Narrative}\\
The average rating given by this user is {\color{blue}4.0} out of 5.\\[0.6em]
\textbf{\# Review History}\\
The user recently commented on the following items:\\[0.4em]
\textbf{Item 1}\\
Item SID:
{\color{blue}
\textless|sid\_begin|\textgreater
\textless s\_a\_60\textgreater
\textless s\_b\_175\textgreater
\textless s\_c\_86\textgreater
\textless s\_d\_158\textgreater
\textless|sid\_end|\textgreater
}\\
Review Title: Did work for me\\
Review Content: I order this cream along with their soap. It actually worked for me but after I finished the tube, I ordered different brand just to get quicker results (BAD IDEA). I am definitely ordering 3 more tubes so that my underarm pigment gets treated completely.\\[0.6em]
\textbf{Item 2}\\
Item SID:
{\color{blue}
\textless|sid\_begin|\textgreater
\textless s\_a\_229\textgreater
\textless s\_b\_165\textgreater
\textless s\_c\_210\textgreater
\textless s\_d\_115\textgreater
\textless|sid\_end|\textgreater
}\\
Review Title: average quality\\
Review Content: This oil has different consistency compare to Josie Maran argan oil. it doesn't worth the money, there is a lot of room for quality improvement.\\[0.6em]
\textbf{Item 3}\\
Item SID:
{\color{blue}
\textless|sid\_begin|\textgreater
\textless s\_a\_23\textgreater
\textless s\_b\_71\textgreater
\textless s\_c\_33\textgreater
\textless s\_d\_5\textgreater
\textless|sid\_end|\textgreater
}\\
Review Title: great product\\
Review Content: it works really well, easy way to get perfect bun. I haven't used the small clip but larger one is really good and long lasting.

\end{tcolorbox}

\begin{tcolorbox}[
    colback=gray!10,
    colframe=gray!60,
    title=Prompt Example for Sequential Preference Modeling,
    fonttitle=\bfseries,
    boxrule=0.5pt,
    rounded corners,
    breakable,
    enhanced
]
\small\ttfamily

\textbf{User ID}: {\color{blue}A1TLDR1V4O48PK}\\[0.6em]
\textbf{\# Purchase History}\\
The user has purchased the following items:\\[0.4em]
\textbf{Item 1}\\
Item SID:
{\color{blue}
\textless|sid\_begin|\textgreater
\textless s\_a\_52\textgreater
\textless s\_b\_72\textgreater
\textless s\_c\_153\textgreater
\textless s\_d\_241\textgreater
\textless|sid\_end|\textgreater
}\\
Title: 120 Color Eyeshadow Palette 3rd Edition\\
Categories: Beauty $>$ Makeup $>$ Eyes $>$ Eye Shadow\\[0.6em]
\textbf{Item 2}\\
Item SID:
{\color{blue}
\textless|sid\_begin|\textgreater
\textless s\_a\_221\textgreater
\textless s\_b\_217\textgreater
\textless s\_c\_124\textgreater
\textless s\_d\_107\textgreater
\textless|sid\_end|\textgreater
}\\
Title: Ion Color Brilliance Brights Semi-Permanent Hair Color Purple\\
Categories: Beauty $>$ Hair Care $>$ Hair Color $>$ Chemical Hair Dyes\\[0.6em]
\textbf{Item 3}\\
Item SID:
{\color{blue}
\textless|sid\_begin|\textgreater
\textless s\_a\_60\textgreater
\textless s\_b\_175\textgreater
\textless s\_c\_86\textgreater
\textless s\_d\_158\textgreater
\textless|sid\_end|\textgreater
}\\
Title: Xtreme Brite Brightening Gel 1oz.\\
Categories: Beauty $>$ Hair Care $>$ Styling Products $>$ Creams, Gels \& Lotions
\end{tcolorbox}

\section{Reasoning Trace Generation with Targeted Sampling}
Here we present two variant of reasoning trace generation with targeted sampling.

\subsection{Reasoning Trace Generation with the Context of Category Hierarchy}
\begin{tcolorbox}[
    colback=gray!10,
    colframe=gray!60,
    title=Prompt Example for Target Sampling with the context of Category Hierarchy,
    fonttitle=\bfseries,
    boxrule=0.5pt,
    rounded corners,
    breakable,
    enhanced
]
\small\ttfamily

\textbf{{\color{sectiongreen}\# System Role}}\\
You are an expert at analyzing e-commerce purchase patterns and predicting user preferences. Given the user's purchase history (with SID identifiers) and a list of candidate items, reason step-by-step about which candidate is most likely to be the user's next purchase.\\[0.8em]

\textbf{{\color{sectiongreen}\# Available Category Hierarchy}}\\
Categories are structured in a hierarchy. Format: \textquotesingle Level0 \textgreater\ Level1 \textgreater\ Level2 \textgreater\ ...\textquotesingle\\[0.4em]
Level 0 (Root): Beauty\\
Level 1 (Main): Bath \& Body, Fragrance, Hair Care, Makeup, Skin Care, ...\\
Level 2 (Sub): Conditioners, Shampoos, Styling Products, Styling Tools, ...\\
Level 3 (Product Type): Creams, Hair Dryers, Irons, Mousses \& Foams, ...\\
Level 4 (Specific): Curling Irons, Flattening Irons, ...\\[0.8em]

\textbf{{\color{sectiongreen}\# User Purchase History}}\\
The user recently purchased the following items:\\[0.4em]

\textbf{Item 1}\\
Item SID: {\color{sidblue}\textless|sid\_begin|\textgreater\textless s\_a\_173\textgreater\textless s\_b\_97\textgreater\textless s\_c\_226\textgreater\textless s\_d\_18\textgreater\textless|sid\_end|\textgreater}\\
Title: L'Oreal Paris EverSleek Sulfate-Free Smoothing System Intense Smoothing Conditioner\\
Categories: Beauty \textgreater\ Hair Care \textgreater\ Conditioners\\[0.4em]

\textbf{Item 2}\\
Item SID: {\color{sidblue}\textless|sid\_begin|\textgreater\textless s\_a\_155\textgreater\textless s\_b\_232\textgreater\textless s\_c\_47\textgreater\textless s\_d\_47\textgreater\textless|sid\_end|\textgreater}\\
Title: Dove Damage Therapy Intensive Repair Daily Super Conditioner\\
Categories: Beauty \textgreater\ Hair Care \textgreater\ Conditioners\\[0.4em]

\textbf{Item 3}\\
Item SID: {\color{sidblue}\textless|sid\_begin|\textgreater\textless s\_a\_173\textgreater\textless s\_b\_79\textgreater\textless s\_c\_173\textgreater\textless s\_d\_249\textgreater\textless|sid\_end|\textgreater}\\
Title: L'Oreal Paris EverStrong Sulfate-Free Fortify System Overnight Hair Repair Treatment\\
Categories: Beauty \textgreater\ Hair Care \textgreater\ Hair \& Scalp Treatments\\[0.4em]

\textbf{Item 4}\\
Item SID: {\color{sidblue}\textless|sid\_begin|\textgreater\textless s\_a\_6\textgreater\textless s\_b\_13\textgreater\textless s\_c\_249\textgreater\textless s\_d\_8\textgreater\textless|sid\_end|\textgreater}\\
Title: Aussie Hair Insurance Leave-In Conditioner\\
Categories: Beauty \textgreater\ Hair Care \textgreater\ Conditioners\\[0.4em]

\textbf{Item 5}\\
Item SID: {\color{sidblue}\textless|sid\_begin|\textgreater\textless s\_a\_27\textgreater\textless s\_b\_159\textgreater\textless s\_c\_86\textgreater\textless s\_d\_25\textgreater\textless|sid\_end|\textgreater}\\
Title: L'Oreal Paris EverSleek Humidity Defying Leave-In Creme\\
Categories: Beauty \textgreater\ Hair Care \textgreater\ Styling Products \textgreater\ Hair Styling Serums\\[0.4em]

{\color{gray}... (Items 6-9 follow same format) ...}\\[0.8em]

\textbf{{\color{sectiongreen}\# Candidate Items}}\\[0.4em]

\textbf{Candidate 1}: Herbal Essences Hello Hydration Conditioner\\
Categories: Beauty \textgreater\ Hair Care \textgreater\ Conditioners\\[0.3em]

\textbf{Candidate 2}: Herbal Essences Tousle Me Softly Conditioner\\
Categories: Beauty \textgreater\ Hair Care \textgreater\ Conditioners\\[0.3em]

\textbf{{Candidate 3}}: Remington Salon Collection Ceramic Hair Straightener\\
Categories: {Beauty \textgreater\ Hair Care \textgreater\ Styling Tools \textgreater\ Irons \textgreater\ Flattening Irons}\\[0.3em]

\textbf{Candidate 4}: Head \& Shoulders Clinical Strength Dandruff Shampoo\\
Categories: Beauty \textgreater\ Hair Care \textgreater\ Shampoos\\[0.3em]

{\color{gray}... (Candidates 5-10 follow same format) ...}\\[0.8em]

\textbf{{\color{sectiongreen}\# Task}}\\
The correct answer is {\color{candidatepurple}Candidate 3} (Remington Ceramic Hair Straightener).\\
Generate a step-by-step reasoning trace explaining why this candidate is the best match.\\[0.8em]

\textbf{{\color{taskred}Critical Guidelines:}}\\
1. \textbf{Cite items by SID}: When referring to purchase history, use their SID directly\\
2. Focus on analyzing patterns in the user's purchase history\\
3. Do NOT use phrases like ``target item'' or ``the target''\\
4. Each step: Reasoning line (with SID citations) + Category line\\[0.8em]

\textbf{{\color{sectiongreen}\# Example Output Format}}\\[0.4em]
Step 1 Reasoning: ``Looking at the purchase history, {\color{sidblue}\textless|sid\_begin|\textgreater...\textless|sid\_end|\textgreater} and {\color{sidblue}\textless|sid\_begin|\textgreater...\textless|sid\_end|\textgreater} are both in the hair care category...''\\
Step 1 Category: ``Beauty \textgreater\ Hair Care''\\[0.3em]
Step 2 Reasoning: ``The recent purchases of {\color{sidblue}\textless|sid\_begin|\textgreater...\textless|sid\_end|\textgreater} (a deep conditioner) suggest...''\\
Step 2 Category: ``Beauty \textgreater\ Hair Care \textgreater\ Conditioners''\\[0.3em]
Step 3 Reasoning: ``Based on their pattern of purchasing complementary hair care products...''\\
Step 3 Category: ``Beauty \textgreater\ Hair Care \textgreater\ Styling Tools \textgreater\ Irons \textgreater\ Flattening Irons''\\[0.8em]

\textbf{{\color{sectiongreen}\# Your Response}}\\
Generate your response following the EXACT format above with all required lines:\\
- Step 1 Reasoning + Step 1 Category (broad)\\
- Step 2 Reasoning + Step 2 Category (specific)\\
- Step 3 Reasoning + Step 3 Category (final: Flattening Irons)

\end{tcolorbox}

%% REASONING TRACE OUTPUT
\begin{tcolorbox}[
    colback=blue!5,
    colframe=blue!60,
    title=Generated Reasoning Trace (Model Output),
    fonttitle=\bfseries,
    boxrule=0.5pt,
    rounded corners,
    breakable,
    enhanced
]
\small\ttfamily\raggedright

\{\par
\hspace{1em}"explanation": "\par\vspace{0.4em}

\hspace{1em}\textbf{{\color{sectiongreen}Step 1 Reasoning}}: ``The user has a history of purchasing hair conditioners and treatments, including {\color{sidblue}<|sid\_begin|><s\_a\_173><s\_b\_97><s\_c\_226><s\_d\_18><|sid\_end|>} (a smoothing conditioner), {\color{sidblue}<|sid\_begin|><s\_a\_173><s\_b\_79><s\_c\_173><s\_d\_249><|sid\_end|}> (an overnight repair treatment), and {\color{sidblue}<|sid\_begin|><s\_a\_49><s\_b\_13><s\_c\_160><s\_d\_43><|sid\_end|>} (a smoothing balm). This indicates a focus on hair health and managing issues like frizz and breakage.''\par\vspace{0.3em}

\hspace{1em}\textbf{{\color{sectiongreen}Step 1 Category}}: {\color{candidatepurple}``Beauty > Hair Care''}\par\vspace{0.5em}

\hspace{1em}\textbf{{\color{sectiongreen}Step 2 Reasoning}}: ``The repeated purchases of conditioners and treatments, such as {\color{sidblue}<|sid\_begin|><s\_a\_6><s\_b\_13><s\_c\_249><s\_d\_8><|sid\_end|>} (a leave-in conditioner) and {\color{sidblue}<|sid\_begin|><s\_a\_148><s\_b\_12><s\_c\_189><s\_d\_62><|sid\_end|>} (a curl-enhancing mousse), suggest the user is invested in maintaining smooth, manageable hair. This progression from basic conditioning to styling products indicates a need for tools to maintain the desired hair texture and style.''\par\vspace{0.3em}

\hspace{1em}Based on the user's purchase history and category patterns, {\color{candidatepurple}Candidate 3} best matches the predicted category and complements their recent purchases.",\par\vspace{0.6em}

\hspace{1em}"recommendations": [{\color{candidatepurple}\textbf{"3"}}, "1", "2", "4", "5", "6", "7", "8", "9", "10"]\par
\}\par

\end{tcolorbox}

\vspace{1em}

%% LEGEND
\begin{tcolorbox}[
    colback=white,
    colframe=gray!40,
    boxrule=0.3pt,
    rounded corners
]
\centering\small
\textbf{Color Legend:}\quad
{\color{sectiongreen}$\blacksquare$} Section Headers \quad
{\color{sidblue}$\blacksquare$} Semantic IDs (SID) \quad
{\color{candidatepurple}$\blacksquare$} Predictions/Answers \quad
{\color{taskred}$\blacksquare$} Guidelines \quad
{\color{gray}$\blacksquare$} Placeholders/Think Tags
\end{tcolorbox}

\subsection{Reasoning Trace Generation without the Context of Category Hierarchy}

\begin{tcolorbox}[
    colback=gray!10,
    colframe=gray!60,
    title=Prompt Example for Target Sampling without the context of Category Hierarchy,
    fonttitle=\bfseries,
    boxrule=0.5pt,
    rounded corners,
    breakable,
    enhanced
]
\small\ttfamily\raggedright

\textbf{{\color{sectiongreen}\# System Role}}\par\vspace{0.3em}
You are an expert at analyzing e-commerce purchase patterns and predicting user preferences. Given the user's purchase history (with SID identifiers) and a list of candidate items, reason step-by-step about which candidate is most likely to be the user's next purchase.\par\vspace{0.8em}

\textbf{{\color{sectiongreen}\# User Purchase History}}\par\vspace{0.3em}

\textbf{Item 1}\par
Item SID: {\color{sidblue}<|sid\_begin|><s\_a\_173><s\_b\_97><s\_c\_226><s\_d\_18><|sid\_end|>}\par
Title: L'Oreal Paris EverSleek Sulfate-Free Smoothing System Intense Smoothing Conditioner\par
Categories: Beauty > Hair Care > Conditioners\par\vspace{0.4em}

\textbf{Item 2}\par
Item SID: {\color{sidblue}<|sid\_begin|><s\_a\_155><s\_b\_232><s\_c\_47><s\_d\_47><|sid\_end|>}\par
Title: Dove Damage Therapy Intensive Repair Daily Super Conditioner\par
Categories: Beauty > Hair Care > Conditioners\par\vspace{0.4em}

\textbf{Item 3}\par
Item SID: {\color{sidblue}<|sid\_begin|><s\_a\_173><s\_b\_79><s\_c\_173><s\_d\_249><|sid\_end|>}\par
Title: L'Oreal Paris EverStrong Sulfate-Free Fortify System Overnight Hair Repair Treatment\par
Categories: Beauty > Hair Care > Hair \& Scalp Treatments\par\vspace{0.4em}

\textbf{Item 4}\par
Item SID: {\color{sidblue}<|sid\_begin|><s\_a\_6><s\_b\_13><s\_c\_249><s\_d\_8><|sid\_end|>}\par
Title: Aussie Hair Insurance Leave-In Conditioner\par
Categories: Beauty > Hair Care > Conditioners\par\vspace{0.4em}

\textbf{Item 5}\par
Item SID: {\color{sidblue}<|sid\_begin|><s\_a\_27><s\_b\_159><s\_c\_86><s\_d\_25><|sid\_end|>}\par
Title: L'Oreal Paris EverSleek Humidity Defying Leave-In Creme\par
Categories: Beauty > Hair Care > Styling Products > Hair Styling Serums\par\vspace{0.4em}

{\color{gray}... (Items 6--9 follow same format) ...}\par\vspace{0.8em}

\textbf{{\color{sectiongreen}\# Candidate Items}}\par\vspace{0.4em}

\textbf{Candidate 1}: Herbal Essences Hello Hydration Conditioner\par
Categories: Beauty > Hair Care > Conditioners\par\vspace{0.3em}

\textbf{Candidate 2}: Herbal Essences Tousle Me Softly Conditioner\par
Categories: Beauty > Hair Care > Conditioners\par\vspace{0.3em}

\textbf{{Candidate 3}}: {Remington Salon Collection Ceramic Hair Straightener}\par
Categories: {Beauty > Hair Care > Styling Tools > Irons > Flattening Irons}\par\vspace{0.3em}

\textbf{Candidate 4}: Head \& Shoulders Clinical Strength Dandruff Shampoo\par
Categories: Beauty > Hair Care > Shampoos\par\vspace{0.3em}

{\color{gray}... (Candidates 5--10 follow same format) ...}\par\vspace{0.8em}

\textbf{{\color{sectiongreen}\# Task}}\par\vspace{0.3em}
The correct answer is {\color{candidatepurple}Candidate 3} (Remington Ceramic Hair Straightener).\par
Generate a step-by-step reasoning trace (3 steps) explaining why this candidate is the best match.\par\vspace{0.8em}

\textbf{{\color{taskred}Critical Guidelines:}}\par\vspace{0.3em}
1. \textbf{Cite items by SID}: When referring to purchase history, use their SID directly\par
2. Focus on analyzing patterns in the user's purchase history\par
3. Do NOT use phrases like ``target item'' or ``the target''\par
4. Think about sequential purchase patterns (e.g., shampoo -> conditioner, phone -> case)\par\vspace{0.8em}

\textbf{{\color{sectiongreen}\# Example Output Format}}\par\vspace{0.4em}
Step 1: ``Looking at the user's purchase history, {\color{sidblue}<|sid\_begin|>...<|sid\_end|>} and {\color{sidblue}<|sid\_begin|>...<|sid\_end|>} are both hair care products in the conditioning category...''\par\vspace{0.3em}
Step 2: ``The recent purchase of {\color{sidblue}<|sid\_begin|>...<|sid\_end|>} (a deep conditioning treatment) suggests the user is looking for products that provide intensive hair care...''\par\vspace{0.3em}
Step 3: ``Based on this pattern, a hydrating hair mask would complement their existing routine...''\par\vspace{0.8em}

\textbf{{\color{sectiongreen}\# Your Response}}\par\vspace{0.3em}
Format your response as:\par
Step 1: ``<analyze the user's purchase patterns, citing specific items by their SID>''\par
Step 2: ``<identify what type of product would complement their recent purchases, citing items by SID>''\par
Step 3: ``<describe the specific product characteristics that make this candidate the best match>''\par

\end{tcolorbox}

\vspace{1em}

%% REASONING TRACE OUTPUT
\begin{tcolorbox}[
    colback=blue!5,
    colframe=blue!60,
    title=Generated Reasoning Trace (Model Output),
    fonttitle=\bfseries,
    boxrule=0.5pt,
    rounded corners,
    breakable,
    enhanced
]
\small\ttfamily\raggedright

\{\par
\hspace{1em}"explanation": "\par\vspace{0.4em}

\hspace{1em}\textbf{{\color{sectiongreen}Step 1}}: ``Looking at the user's purchase history, {\color{sidblue}<|sid\_begin|><s\_a\_173><s\_b\_97><s\_c\_226><s\_d\_18><|sid\_end|>}, {\color{sidblue}<|sid\_begin|><s\_a\_173><s\_b\_79><s\_c\_173><s\_d\_249><|sid\_end|>}, and {\color{sidblue}<|sid\_begin|><s\_a\_27><s\_b\_159><s\_c\_86><s\_d\_25><|sid\_end|>} are all smoothing and frizz control products, indicating a strong interest in managing frizz and achieving smooth hair.''\par\vspace{0.4em}

\hspace{1em}\textbf{{\color{sectiongreen}Step 2}}: ``The user has consistently purchased hair conditioners and treatments, as seen with {\color{sidblue}<|sid\_begin|><s\_a\_155><s\_b\_232><s\_c\_47><s\_d\_47><|sid\_end|>}, {\color{sidblue}<|sid\_begin|><s\_a\_6><s\_b\_13><s\_c\_249><s\_d\_8><|sid\_end|>}, {\color{sidblue}<|sid\_begin|><s\_a\_6><s\_b\_134><s\_c\_71><s\_d\_245><|sid\_end|>}, and {\color{sidblue}<|sid\_begin|><s\_a\_6><s\_b\_68><s\_c\_105><s\_d\_181><|sid\_end|>}, suggesting they prioritize conditioning and deep hair care.''\par\vspace{0.4em}

\hspace{1em}\textbf{{\color{sectiongreen}Step 3}}: ``{\color{candidatepurple}Candidate 3} is a ceramic hair straightener with wide plates, which aligns with the user's focus on smoothing and frizz control. It complements their existing smoothing conditioners and treatments by providing a styling tool that enhances smoothness, making it the best match for their hair care routine.''\par\vspace{0.4em}

\hspace{1em}Based on the user's purchase history and category patterns, {\color{candidatepurple}Candidate 3} best matches the predicted category and complements their recent purchases.",\par\vspace{0.6em}

\hspace{1em}"recommendations": [{\color{candidatepurple}\textbf{"3"}}, "1", "2", "4", "5", "6", "7", "8", "9", "10"]\par
\}\par

\end{tcolorbox}

\vspace{1em}

%% LEGEND
\begin{tcolorbox}[
    colback=white,
    colframe=gray!40,
    boxrule=0.3pt,
    rounded corners
]
\centering\small
\textbf{Color Legend:}\quad
{\color{sectiongreen}$\blacksquare$} Section Headers / Steps \quad
{\color{sidblue}$\blacksquare$} Semantic IDs (SID) \quad
{\color{candidatepurple}$\blacksquare$} Predictions / Correct Answer \quad
{\color{taskred}$\blacksquare$} Guidelines \quad
{\color{gray}$\blacksquare$} Placeholders
\end{tcolorbox}

\section{Reasoning Trace Generation with Rejection Sampling}
\subsection{Reasoning Trace Generation with the Context of Category Hierarchy}

\begin{tcolorbox}[
    colback=gray!10,
    colframe=gray!60,
    title=Prompt Example for Rejection Sampling with the context of Category Hierarchy,
    fonttitle=\bfseries,
    boxrule=0.5pt,
    rounded corners,
    breakable,
    enhanced
]
\small\ttfamily\raggedright

\textbf{{\color{sectiongreen}\# System Role}}\par\vspace{0.3em}
You are an expert at analyzing e-commerce purchase patterns and predicting user preferences.\par\vspace{0.3em}
Given the user's purchase history (with SID identifiers) and a list of candidate items, you need to predict which candidate is MOST LIKELY to be the user's next purchase.\par\vspace{0.8em}

\textbf{{\color{sectiongreen}\# Available Category Hierarchy}}\par\vspace{0.3em}
Categories are structured in a hierarchy. Format: 'Level0 > Level1 > Level2 > ...'\par\vspace{0.3em}
Level 0 (Root): Beauty\par
Level 1 (Main): Bath \& Body, Fragrance, Hair Care, Makeup, Skin Care, Tools \& Accessories\par
Level 2 (Sub): Conditioners, Shampoos, Styling Products, Hair \& Scalp Treatments, ...\par
Level 3 (Product Type): Creams, Gels \& Lotions, Hair Sprays, Oils \& Serums, ...\par
Level 4 (Specific): Curling Irons, Flattening Irons, ...\par
Level 5 (Variant): Retinol, Glycolic Acid, ...\par\vspace{0.8em}

\textbf{{\color{sectiongreen}\# User Purchase History}}\par\vspace{0.3em}
The user recently purchased the following items:\par\vspace{0.4em}

\textbf{Item 1}\par
Item SID: {\color{sidblue}<|sid\_begin|><s\_a\_57><s\_b\_7><s\_c\_213><s\_d\_26><|sid\_end|>}\par
Title: One'n Only Argan Oil Leave-In Treatment\par
Categories: Beauty > Hair Care > Hair \& Scalp Treatments\par\vspace{0.4em}

\textbf{Item 2}\par
Item SID: {\color{sidblue}<|sid\_begin|><s\_a\_7><s\_b\_112><s\_c\_18><s\_d\_204><|sid\_end|>}\par
Title: Hair One Cleanser and Conditioner with Argan Oil for Curly Hair 12 oz\par
Categories: Beauty > Hair Care > Shampoos\par\vspace{0.8em}

\textbf{{\color{sectiongreen}\# Candidate Items}}\par\vspace{0.4em}

\textbf{Candidate 1}: One 'n Only Argan Oil Styling Cream, 10 fl. oz.\par
Categories: Beauty > Hair Care > Styling Products > Creams, Gels \& Lotions\par\vspace{0.3em}

\textbf{Candidate 2}: One 'n Only Argan Oil Spray Treatment 6 fl. oz\par
Categories: Beauty > Hair Care > Hair \& Scalp Treatments\par\vspace{0.3em}

\textbf{Candidate 3}: Deva Devacurl One Condition Conditioner, 12 Ounce\par
Categories: Beauty > Hair Care > Conditioners\par\vspace{0.3em}

\textbf{Candidate 4}: Giovanni Hair Care - Direct Leave-In Conditioner, 8.5 fl oz\par
Categories: Beauty > Hair Care > Hair Relaxers > Conditioners\par\vspace{0.3em}

\textbf{Candidate 5}: Paul Mitchell The Conditioner, Leave-in Moisturizer, 10.14-ounce\par
Categories: Beauty > Hair Care > Conditioners\par\vspace{0.3em}

\textbf{Candidate 6}: Kinky-Curly Knot Today Leave In Conditioner/Detangler - 8 oz\par
Categories: Beauty > Hair Care > Conditioners\par\vspace{0.3em}

\textbf{Candidate 7}: Hair One Cleanser and Conditioner with Olive Oil for Dry Hair 12 oz\par
Categories: Beauty > Hair Care > Hair Loss Products > Conditioners\par\vspace{0.3em}

\textbf{Candidate 8}: It's A 10 Miracle Leave In Product, 4-Ounces\par
Categories: Beauty > Hair Care > Conditioners\par\vspace{0.3em}

\textbf{Candidate 9}: It's A 10 Miracle Moisture Shampoo, 10-Ounce Bottle\par
Categories: Beauty > Hair Care > Shampoos\par\vspace{0.3em}

\textbf{Candidate 10}: Dabur Vatika Enriched Coconut Hair Oil 150ml (Pack of 2)\par
Categories: Beauty > Hair Care > Hair \& Scalp Treatments\par\vspace{0.8em}

\textbf{{\color{sectiongreen}\# Task}}\par\vspace{0.3em}
Analyze the user's purchase history and predict which candidate they are most likely to purchase next.\par\vspace{0.8em}

\textbf{{\color{taskred}Critical Guidelines:}}\par\vspace{0.3em}
1. \textbf{Cite items by SID}: When referring to items in the purchase history, cite them directly using their SID (e.g., {\color{sidblue}<|sid\_begin|><s\_a\_99><s\_b\_19><s\_c\_220><s\_d\_204><|sid\_end|>}). This allows the model to learn to reason with semantic IDs naturally interleaved with language.\par
2. Focus on analyzing patterns in the user's purchase history\par
3. Each step should have both a Reasoning line (free-form with SID citations) AND a Category line\par\vspace{0.8em}

\textbf{{\color{sectiongreen}\# Example Output Format}}\par\vspace{0.4em}
Step 1 Reasoning: ``Looking at the purchase history, {\color{sidblue}<|sid\_begin|>...<|sid\_end|>} and {\color{sidblue}<|sid\_begin|>...<|sid\_end|>} are both in the hair care category, indicating the user focuses on personal care.''\par
Step 1 Category: ``Beauty > Hair Care''\par\vspace{0.3em}
Step 2 Reasoning: ``The recent purchases of {\color{sidblue}<|sid\_begin|>...<|sid\_end|>} (a deep conditioner) and {\color{sidblue}<|sid\_begin|>...<|sid\_end|>} (a leave-in treatment) suggest the user is building a comprehensive hair care routine focused on moisture and repair.''\par
Step 2 Category: ``Beauty > Hair Care > Conditioners''\par\vspace{0.3em}
Step 3 Reasoning: ``Based on their pattern, Candidate 3 (a hydrating hair mask) would naturally complement their existing routine.''\par
Step 3 Category: ``Beauty > Hair Care > Conditioners''\par
Prediction: Candidate 3\par\vspace{0.8em}

\textbf{{\color{sectiongreen}\# Your Response}}\par\vspace{0.3em}
Generate your response following the EXACT format above. You MUST include:\par
-- Step 1 Reasoning (free-form analysis citing items by SID) + Step 1 Category (broad)\par
-- Step 2 Reasoning (narrowing down, citing specific items by SID) + Step 2 Category (more specific)\par
-- Step 3 Reasoning (final prediction explanation) + Step 3 Category (final specific)\par
-- Prediction: Candidate <number>\par\vspace{0.3em}
Reasoning:

\end{tcolorbox}

\vspace{1em}

%% REASONING TRACE OUTPUT
\begin{tcolorbox}[
    colback=blue!5,
    colframe=blue!60,
    title=Generated Reasoning Trace (Model Output),
    fonttitle=\bfseries,
    boxrule=0.5pt,
    rounded corners,
    breakable,
    enhanced
]
\small\ttfamily\raggedright

\{\par
\hspace{1em}"explanation": "\par\vspace{0.4em}

\hspace{1em}\textbf{{\color{sectiongreen}Step 1 Reasoning}}: ``The user has purchased {\color{sidblue}<|sid\_begin|><s\_a\_57><s\_b\_7><s\_c\_213><s\_d\_26><|sid\_end|>} (a leave-in treatment) and {\color{sidblue}<|sid\_begin|><s\_a\_7><s\_b\_112><s\_c\_18><s\_d\_204><|sid\_end|>} (a cleanser and conditioner), both of which are in the hair care category, indicating a focus on hair care products.''\par\vspace{0.3em}

\hspace{1em}\textbf{{\color{sectiongreen}Step 1 Category}}: {\color{candidatepurple}``Beauty > Hair Care''}\par\vspace{0.5em}

\hspace{1em}\textbf{{\color{sectiongreen}Step 2 Reasoning}}: ``The user has already purchased a cleanser and conditioner with argan oil, suggesting a preference for argan oil-based products. The leave-in treatment is also in the hair and scalp treatments category, indicating a focus on treatments. Candidate 1 and Candidate 2 are in the same category as the leave-in treatment, while Candidates 3--10 are in the conditioners or other categories. Candidate 2 is in the hair and scalp treatments category, similar to the user's previous purchase of {\color{sidblue}<|sid\_begin|><s\_a\_57><s\_b\_7><s\_c\_213><s\_d\_26><|sid\_end|>}.''\par\vspace{0.5em}

\hspace{1em}Based on the user's purchase history and category patterns, {\color{candidatepurple}Candidate 1} best matches the predicted category and complements their recent purchases.",\par\vspace{0.6em}

\hspace{1em}"recommendations": [{\color{candidatepurple}\textbf{"1"}}, "2", "3", "4", "5", "6", "7", "8", "9", "10"]\par
\}\par

\end{tcolorbox}

\vspace{1em}

%% LEGEND
\begin{tcolorbox}[
    colback=white,
    colframe=gray!40,
    boxrule=0.3pt,
    rounded corners
]
\centering\small
\textbf{Color Legend:}\quad
{\color{sectiongreen}$\blacksquare$} Section Headers \quad
{\color{sidblue}$\blacksquare$} Semantic IDs (SID) \quad
{\color{candidatepurple}$\blacksquare$} Predictions/Answers \quad
{\color{taskred}$\blacksquare$} Guidelines \quad
{\color{gray}$\blacksquare$} Placeholders/Think Tags
\end{tcolorbox}

\subsection{Reasoning Trace Generation without the Context of Category Hierarchy}

\begin{tcolorbox}[
    colback=gray!10,
    colframe=gray!60,
    title=Prompt Example for E-Commerce Purchase Prediction (Simple Format),
    fonttitle=\bfseries,
    boxrule=0.5pt,
    rounded corners,
    breakable,
    enhanced
]
\small\ttfamily\raggedright

\textbf{{\color{sectiongreen}\# System Role}}\par\vspace{0.3em}
You are an expert at analyzing e-commerce purchase patterns and predicting user preferences.\par\vspace{0.3em}
Given the user's purchase history (with SID identifiers) and a list of candidate items, you need to predict which candidate is MOST LIKELY to be the user's next purchase.\par\vspace{0.8em}

\textbf{{\color{sectiongreen}\# User Purchase History}}\par\vspace{0.3em}
The user recently purchased the following items:\par\vspace{0.4em}

\textbf{Item 1}\par
Item SID: {\color{sidblue}<|sid\_begin|><s\_a\_57><s\_b\_7><s\_c\_213><s\_d\_26><|sid\_end|>}\par
Title: One'n Only Argan Oil Leave-In Treatment\par
Categories: Beauty > Hair Care > Hair \& Scalp Treatments\par\vspace{0.4em}

\textbf{Item 2}\par
Item SID: {\color{sidblue}<|sid\_begin|><s\_a\_7><s\_b\_112><s\_c\_18><s\_d\_204><|sid\_end|>}\par
Title: Hair One Cleanser and Conditioner with Argan Oil for Curly Hair 12 oz\par
Categories: Beauty > Hair Care > Shampoos\par\vspace{0.8em}

\textbf{{\color{sectiongreen}\# Candidate Items}}\par\vspace{0.4em}

\textbf{Candidate 1}: One 'n Only Argan Oil Styling Cream, 10 fl. oz.\par
Categories: Beauty > Hair Care > Styling Products > Creams, Gels \& Lotions\par\vspace{0.3em}

\textbf{Candidate 2}: One 'n Only Argan Oil Spray Treatment 6 fl. oz\par
Categories: Beauty > Hair Care > Hair \& Scalp Treatments\par\vspace{0.3em}

\textbf{Candidate 3}: Deva Devacurl One Condition Conditioner, 12 Ounce\par
Categories: Beauty > Hair Care > Conditioners\par\vspace{0.3em}

\textbf{Candidate 4}: Giovanni Hair Care - Direct Leave-In Conditioner, 8.5 fl oz\par
Categories: Beauty > Hair Care > Hair Relaxers > Conditioners\par\vspace{0.3em}

\textbf{Candidate 5}: Paul Mitchell The Conditioner, Leave-in Moisturizer, 10.14-ounce\par
Categories: Beauty > Hair Care > Conditioners\par\vspace{0.3em}

\textbf{Candidate 6}: Kinky-Curly Knot Today Leave In Conditioner/Detangler - 8 oz\par
Categories: Beauty > Hair Care > Conditioners\par\vspace{0.3em}

\textbf{Candidate 7}: Hair One Cleanser and Conditioner with Olive Oil for Dry Hair 12 oz\par
Categories: Beauty > Hair Care > Hair Loss Products > Conditioners\par\vspace{0.3em}

\textbf{Candidate 8}: It's A 10 Miracle Leave In Product, 4-Ounces\par
Categories: Beauty > Hair Care > Conditioners\par\vspace{0.3em}

\textbf{Candidate 9}: It's A 10 Miracle Moisture Shampoo, 10-Ounce Bottle\par
Categories: Beauty > Hair Care > Shampoos\par\vspace{0.3em}

\textbf{Candidate 10}: Dabur Vatika Enriched Coconut Hair Oil 150ml (Pack of 2)\par
Categories: Beauty > Hair Care > Hair \& Scalp Treatments\par\vspace{0.8em}

\textbf{{\color{sectiongreen}\# Task}}\par\vspace{0.3em}
Analyze the user's purchase history and predict which candidate they are most likely to purchase next.\par
Think step-by-step about the patterns and preferences shown in their history.\par\vspace{0.8em}

\textbf{{\color{taskred}Critical Guidelines:}}\par\vspace{0.3em}
1. \textbf{Cite items by SID}: When referring to items in the purchase history, cite them directly using their SID (e.g., {\color{sidblue}<|sid\_begin|><s\_a\_99><s\_b\_19><s\_c\_220><s\_d\_204><|sid\_end|>}). This allows the model to learn to reason with semantic IDs naturally interleaved with language.\par
2. Focus on analyzing patterns in the user's purchase history\par
3. Think about sequential purchase patterns (e.g., shampoo $\rightarrow$ conditioner, phone $\rightarrow$ case)\par\vspace{0.8em}

\textbf{{\color{sectiongreen}\# Example Output Format}}\par\vspace{0.4em}
Step 1: ``Looking at the user's purchase history, {\color{sidblue}<|sid\_begin|>...<|sid\_end|>} and {\color{sidblue}<|sid\_begin|>...<|sid\_end|>} are both hair care products, indicating a focus on hair maintenance.''\par\vspace{0.3em}
Step 2: ``The recent purchase of {\color{sidblue}<|sid\_begin|>...<|sid\_end|>} (a deep conditioning treatment) suggests the user is looking for products that provide intensive hair care.''\par\vspace{0.3em}
Step 3: ``Based on this pattern, Candidate 3 would be the natural next purchase as it complements their existing routine.''\par
Prediction: Candidate 3\par\vspace{0.8em}

\textbf{{\color{sectiongreen}\# Your Response Format}}\par\vspace{0.3em}
Step 1: ``<analyze the user's purchase patterns, citing specific items by their SID>''\par
Step 2: ``<identify what type of product would complement their recent purchases, citing items by SID>''\par
Step 3: ``<explain which candidate best matches the user's needs>''\par
Prediction: Candidate <number>\par\vspace{0.3em}
Reasoning:

\end{tcolorbox}

\vspace{1em}

%% REASONING TRACE OUTPUT
\begin{tcolorbox}[
    colback=blue!5,
    colframe=blue!60,
    title=Generated Reasoning Trace (Model Output),
    fonttitle=\bfseries,
    boxrule=0.5pt,
    rounded corners,
    breakable,
    enhanced
]
\small\ttfamily\raggedright

\{\par
\hspace{1em}"explanation": "\par\vspace{0.4em}

\hspace{1em}\textbf{{\color{sectiongreen}Step 1}}: ``Looking at the user's purchase history, {\color{sidblue}<|sid\_begin|><s\_a\_57><s\_b\_7><s\_c\_213><s\_d\_26><|sid\_end|>} and {\color{sidblue}<|sid\_begin|><s\_a\_7><s\_b\_112><s\_c\_18><s\_d\_204><|sid\_end|>} are both hair care products, specifically for curly hair, indicating a focus on maintaining curly hair with argan oil.''\par\vspace{0.5em}

\hspace{1em}\textbf{{\color{sectiongreen}Step 2}}: ``The purchase of a cleanser and conditioner ({\color{sidblue}<|sid\_begin|><s\_a\_7><s\_b\_112><s\_c\_18><s\_d\_204><|sid\_end|>}) suggests the user follows a routine of shampoo and conditioner, which is typically followed by styling or treatment products.''\par\vspace{0.5em}

\hspace{1em}\textbf{{\color{sectiongreen}Step 3}}: ``Candidate 1 (One 'n Only Argan Oil Styling Cream) and Candidate 2 (One 'n Only Argan Oil Spray Treatment) both align with the brand and product type the user has previously purchased. However, since styling products are typically used after cleansing and conditioning, {\color{candidatepurple}Candidate 1} is the most logical next step in the user's hair care routine.''\par\vspace{0.5em}

\hspace{1em}Based on the user's purchase history and category patterns, {\color{candidatepurple}Candidate 1} best matches the predicted category and complements their recent purchases.",\par\vspace{0.6em}

\hspace{1em}"recommendations": [{\color{candidatepurple}\textbf{"1"}}, "2", "3", "4", "5", "6", "7", "8", "9", "10"]\par
\}\par

\end{tcolorbox}

\vspace{1em}

%% LEGEND
\begin{tcolorbox}[
    colback=white,
    colframe=gray!40,
    boxrule=0.3pt,
    rounded corners
]
\centering\small
\textbf{Color Legend:}\quad
{\color{sectiongreen}$\blacksquare$} Section Headers \quad
{\color{sidblue}$\blacksquare$} Semantic IDs (SID) \quad
{\color{candidatepurple}$\blacksquare$} Predictions/Answers \quad
{\color{taskred}$\blacksquare$} Guidelines \quad
{\color{gray}$\blacksquare$} Placeholders
\end{tcolorbox}

\section{Methodology Details}
\label{sec:appendix method}

\begin{tcolorbox}[
    colback=gray!8,
    colframe=gray!60,
    title=\textbf{Robust Parsing of Reasoning-aware Re-ranking Output},
    fonttitle=\bfseries,
    boxrule=0.5pt,
    rounded corners,
    breakable,
    enhanced
]

\small
\textbf{Input:} raw LLM output $O$, number of candidates $|D|$ \\
\textbf{Output:} reasoning trace $O_R$ (optional), validated ranking $O_D$ (or \texttt{None})

\vspace{0.4em}

\hrule
\vspace{0.4em}

\textbf{Algorithm: Robust Re-ranking Output Parsing}

\begin{enumerate}
    \item \textbf{Initialize} $O_R \leftarrow \texttt{None}$, $O_D \leftarrow \texttt{None}$

    \vspace{0.3em}

    \item \textbf{Stage I: JSON-based Structured Parsing}
    \begin{enumerate}
        \item Extract all JSON-like substrings from $O$
        \item Traverse substrings in reverse order
        \item For each substring:
        \begin{itemize}
            \item Attempt JSON parsing
            \item If successful:
            \begin{itemize}
                \item Extract reasoning field (\texttt{explanation} or \texttt{reasoning})
                \item Extract ranking field (\texttt{recommendations} or \texttt{ranking})
                \item If ranking is a non-empty list, break
            \end{itemize}
        \end{itemize}
    \end{enumerate}

    {\color{gray}\textit{$\triangleright$ Prefer well-structured outputs while tolerating extra text}}

    \vspace{0.5em}

    \item \textbf{Stage II: Regex-based Fallback Parsing (if Stage I fails)}
    \begin{enumerate}
        \item Search for ranking patterns using regular expressions
        \item Extract all numeric indices from the matched span
        \item Optionally extract reasoning text if present
    \end{enumerate}

    {\color{gray}\textit{$\triangleright$ Handles malformed or partially structured LLM generations}}

    \vspace{0.5em}

    \item \textbf{Stage III: Ranking Validation and Canonicalization}
    \begin{enumerate}
        \item Filter ranking entries to valid integers in $[1, |D|]$
        \item Remove duplicates while preserving order
        \item If no valid index remains, return $(\texttt{None}, \texttt{None})$
        \item Convert to zero-based indices
        \item Append missing candidates to form a full permutation of length $|D|$
    \end{enumerate}

    {\color{gray}\textit{$\triangleright$ Guarantees a valid permutation for reward computation}}
\end{enumerate}

\vspace{0.4em}
\hrule
\vspace{0.4em}

\textbf{Return:} $(O_R, O_D)$ if a valid ranking is recovered; otherwise $(\texttt{None}, \texttt{None})$

\end{tcolorbox}

\section{Case Study}
\label{sec:appendix case study}

\begin{tcolorbox}[
    colback=gray!10,
    colframe=gray!60,
    title=GR2's Case Study for Targeted Reasoning with Knowledge Priming,
    fonttitle=\bfseries,
    boxrule=0.5pt,
    rounded corners,
    breakable,
    enhanced
]
\small\ttfamily\raggedright

\textbf{{\color{sectiongreen}\# System Role}}\par\vspace{0.3em}
You are an expert at analyzing e-commerce purchase patterns and predicting user preferences. Given the user's purchase history (with SID identifiers) and a list of candidate items, predict which item the user is most likely to purchase next by re-ranking the pre-ranked candidates.\par\vspace{0.8em}

\textbf{{\color{sectiongreen}\# User Purchase History}}\par\vspace{0.3em}

\textbf{Item 1}\par
Item SID: {\color{sidblue}<|sid\_begin|><s\_a\_134><s\_b\_194><s\_c\_131><s\_d\_121><|sid\_end|>}\par
Title: Pureology Hydrate Shampoo 33.8 oz \& Condition 33.8 oz Duo Set\par
Categories: Beauty > Hair Care > Shampoo \& Conditioner Sets\par\vspace{0.4em}

\textbf{Item 2}\par
Item SID: {\color{sidblue}<|sid\_begin|><s\_a\_120><s\_b\_95><s\_c\_240><s\_d\_161><|sid\_end|>}\par
Title: Pureology Hydrate Shampoo, 8.5 Ounce\par
Categories: Beauty > Hair Care > Shampoos\par\vspace{0.4em}

\textbf{Item 3}\par
Item SID: {\color{sidblue}<|sid\_begin|><s\_a\_238><s\_b\_79><s\_c\_36><s\_d\_141><|sid\_end|>}\par
Title: Pureology Anti-Fade Complex Hydrate Conditioner, 8.5 Ounce\par
Categories: Beauty > Hair Care > Conditioners\par\vspace{0.8em}

\textbf{{\color{sectiongreen}\# Candidate Items}}\par\vspace{0.4em}

\textbf{Candidate 1}\par
Item SID: {\color{sidblue}<|sid\_begin|><s\_a\_238><s\_b\_194><s\_c\_120><s\_d\_24><|sid\_end|>}\par
Title: Pravana Pure Light Sulfate-free Brightening Shampoo\par
Categories: Beauty > Hair Care > Shampoos\par\vspace{0.3em}

\textbf{Candidate 2}\par
Item SID: {\color{sidblue}<|sid\_begin|><s\_a\_134><s\_b\_194><s\_c\_59><s\_d\_74><|sid\_end|>}\par
Title: Pureology Hydrate Shampoo 8.5oz and Hydrate Conditioner 8.5oz Duo\par
Categories: Beauty > Hair Care > Shampoo \& Conditioner Sets\par\vspace{0.3em}

\textbf{Candidate 3}\par
Item SID: {\color{sidblue}<|sid\_begin|><s\_a\_155><s\_b\_123><s\_c\_248><s\_d\_251><|sid\_end|>}\par
Title: L'Oreal Paris EverSleek Sulfate-Free Smoothing Shampoo\par
Categories: Beauty > Hair Care > Shampoos\par\vspace{0.3em}

\textbf{Candidate 4}\par
Item SID: {\color{sidblue}<|sid\_begin|><s\_a\_137><s\_b\_63><s\_c\_202><s\_d\_132><|sid\_end|>}\par
Title: Revlon RV544PKF Ionic Ceramic Hair Dryer\par
Categories: Beauty > Hair Care > Styling Tools > Hair Dryers\par\vspace{0.3em}

\textbf{Candidate 5}\par
Item SID: {\color{sidblue}<|sid\_begin|><s\_a\_94><s\_b\_56><s\_c\_151><s\_d\_226><|sid\_end|>}\par
Title: Cetaphil Moisturizing Cream (Pack of 3)\par
Categories: Beauty > Skin Care > Body > Moisturizers\par\vspace{0.3em}

\textbf{Candidate 6}\par
Item SID: {\color{sidblue}<|sid\_begin|><s\_a\_238><s\_b\_178><s\_c\_125><s\_d\_87><|sid\_end|>}\par
Title: Clairol Shimmer Lights Shampoo Blonde \& Silver 31.5oz\par
Categories: Beauty > Hair Care > Shampoos\par\vspace{0.3em}

\textbf{Candidate 7}\par
Item SID: {\color{sidblue}<|sid\_begin|><s\_a\_238><s\_b\_56><s\_c\_125><s\_d\_9><|sid\_end|>}\par
Title: Clairol Professional Shimmer Lights Shampoo 31.5oz\par
Categories: Beauty > Hair Care > Shampoos\par\vspace{0.3em}

\textbf{Candidate 8}\par
Item SID: {\color{sidblue}<|sid\_begin|><s\_a\_134><s\_b\_209><s\_c\_245><s\_d\_42><|sid\_end|>}\par
Title: Agadir Argan Oil Shampoo + Conditioner Combo Set\par
Categories: Beauty > Hair Care > Shampoo \& Conditioner Sets\par\vspace{0.3em}

\textbf{Candidate 9}\par
Item SID: {\color{sidblue}<|sid\_begin|><s\_a\_243><s\_b\_182><s\_c\_89><s\_d\_5><|sid\_end|>}\par
Title: Cricket Technique \#450 Thermal Brush\par
Categories: Beauty > Hair Care > Styling Tools > Brushes\par\vspace{0.3em}

\textbf{Candidate 10}\par
Item SID: {\color{sidblue}<|sid\_begin|><s\_a\_155><s\_b\_202><s\_c\_218><s\_d\_11><|sid\_end|>}\par
Title: Fekkai Glossing Hair Products Starter Kit\par
Categories: Beauty > Hair Care\par\vspace{0.8em}

\textbf{{\color{sectiongreen}\# Model Output}}\par\vspace{0.4em}

\textbf{Reasoning Trace}\par
Step 1 Reasoning: ``The user has a history of purchasing hair care products, specifically shampoo and conditioner sets.  is a Pureology Hydrate Shampoo and Conditioner Duo Set, and  and  are individual Pureology Hydrate Shampoo and Conditioner products. This indicates a consistent interest in hair care, particularly in hydrating products.''\par\vspace{0.3em}
Step 1 Category: ``Beauty > Hair Care''\par\vspace{0.3em}
Step 2 Reasoning: ``The user has specifically purchased hydrating shampoo and conditioner sets, which suggests a focus on maintaining hair health and moisture. The Pureology Hydrate line is known for its hydrating properties, and the user has bought both the shampoo and conditioner, indicating a preference for complete sets. This pattern aligns with the candidate items that are also shampoo and conditioner sets, such as {\color{sidblue}<|sid\_begin|><s\_a\_134><s\_b\_194><s\_c\_59><s\_d\_74><|sid\_end|>} (Candidate 2) and {\color{sidblue}<|sid\_begin|><s\_a\_134><s\_b\_209><s\_c\_245><s\_d\_42><|sid\_end|>} (Candidate 8).''\par\vspace{0.3em}
Based on the user's purchase history and category patterns, Candidate 2 best matches the predicted category and complements their recent purchases.\par\vspace{0.6em}

\textbf{Re-ranked Candidates}\par
[2,\;8,\;1,\;3,\;4,\;5,\;6,\;7,\;9,\;10]\par\vspace{0.8em}

\textbf{{\color{sectiongreen}\# Ground Truth}}\par\vspace{0.4em}
{\color{sidblue}<|sid\_begin|><s\_a\_134><s\_b\_194><s\_c\_59><s\_d\_74><|sid\_end|>} (Candidate 2)

\end{tcolorbox}

\section{Evaluation Metrics}
\label{sec:appendix metrics}

\paragraph{Recall@K.}
Let $\mathcal{U}$ denote the set of users, $i_u^{\ast}$ the ground-truth next item for user $u \in \mathcal{U}$, and $\mathcal{R}_u^{K}$ the set of top-$K$ items ranked by the model.
Recall@K is defined as
\begin{equation}
\mathrm{Recall@}K
= \mathbb E_{u\in\mathcal U}\left[
\mathbb{I}\!\left( i_u^{\ast} \in \mathcal{R}_u^{K} \right)\right],
\end{equation}
where $\mathbb{I}(\cdot)$ is the indicator function.

\paragraph{NDCG@K.}
For user $u$, the Discounted Cumulative Gain at $K$ (DCG@K) is computed as
\begin{equation}
\mathrm{DCG@}K
= \sum_{j=1}^{K}
\frac{2^{\mathrm{rel}_{u,j}} - 1}{\log_2(j+1)},
\end{equation}
where $\mathrm{rel}_{u,j} \in \{0,1\}$ indicates whether the item ranked at position $j$ matches the ground-truth next item $i_u^{\ast}$.

The Ideal DCG at $K$ (IDCG@K) is obtained by placing the ground-truth item at the first position, in the standard leave-one-out evaluation protocol for sequential recommendation which is always $1$: $\mathrm{IDCG@}K
= \frac{1}{\log_2(1+1)} = 1$. NDCG@K is then defined as
\begin{equation}
\mathrm{NDCG@}K
= \mathbb E_{u\in\mathcal U}\left[
\frac{\mathrm{DCG@}K}{\mathrm{IDCG@}K}\right].
\end{equation}

\paragraph{SID Uniqueness/Collision Rate}

To evaluate the quality of the learned SIDs, we measure the \emph{collision rate}, which quantifies how often multiple items are assigned the same SID.
A lower collision rate indicates stronger discriminative power of the tokenizer and a more faithful semantic encoding of items.

Let $\mathcal{I}$ denote the set of items, and let $\mathrm{SID}(i)$ be the SID assigned to item $i \in \mathcal{I}$.
We define the collision set as
\begin{equation}
\mathcal{C}
= \left\{ i \in \mathcal{I} \;\middle|\;
\exists j \in \mathcal{I},\; j \neq i,\;
\mathrm{SID}(i) = \mathrm{SID}(j)
\right\}.
\end{equation}

The \emph{collision rate} is then computed as
\begin{equation}
\mathrm{CollisionRate}
= |\mathcal{C}|/|\mathcal{I}|.
\end{equation}

and the SID uniqueness is computed as
\begin{align}
    \mathrm{Uniqueness} = 1- \mathrm{CollisionRate}
\end{align}

\end{document}